\documentclass[preprint]{aastex62}  

\begin{document}

\title{New Moons of Uranus and Neptune from Ultra-Deep Pencil Beam Surveys}

\author[0000-0003-3145-8682]{Scott S. Sheppard} 
\affil{Earth and Planets Laboratory, Carnegie Institution for Science, 5241 Broad Branch Rd. NW, Washington, DC 20015, USA, ssheppard@carnegiescience.edu}

\author[0000-0003-0773-1888]{David J. Tholen} 
\affil{Institute for Astronomy, University of Hawai'i, Honolulu, HI 96822, USA}

\author[0009-0007-3493-2139]{Marina Brozovic}
\affil{Jet Propulsion Laboratory, California Institute of Technology, 4800 Oak Grove Drive, Pasadena, CA 91109, USA}

\author{Robert Jacobson}
\affil{Jet Propulsion Laboratory, California Institute of Technology, 4800 Oak Grove Drive, Pasadena, CA 91109, USA}

\author[0000-0001-9859-0894]{Chadwick A. Trujillo}
\affil{Department of Astronomy and Planetary Science, Northern Arizona University, Flagstaff, AZ 86011, USA}

\author[0000-0003-0926-2448]{Patryk Sofia Lykawka}
\affil{Kindai University, Shinkamikosaka 228-3, Higashiosaka, Osaka, 577-0813, Japan}

\author[0000-0003-4143-8589]{Mike Alexandersen}
\affil{Center for Astrophysics $|$ Harvard \& Smithsonian, 60 Garden Street, Cambridge, MA 02138, USA}

\begin{abstract}  

We have conducted extremely ultra-deep pencil beam observations for
new satellites around both Uranus and Neptune. Tens of images on
several different nights in 2021, 2022 and 2023 were obtained and
shifted and added together to reach as faint as 26.9 and 27.2
magnitudes in the r-band around Uranus and Neptune, respectively. One
new moon of Uranus, S/2023 U1, and two new moons of Neptune, S/2021 N1
and S/2002 N5, were found. S/2023 U1 was 26.6 mags, is about 7 km in
diameter and has a distant, eccentric and inclined retrograde orbit
similar to Caliban and Stephano, implying these satellites are
fragments from a once larger parent satellite. S/2021 N1 was 26.9
mags, about 14 km in size and has a retrograde orbit similar to Neso
and Psamathe, indicating they are a dynamical family. We find S/2021
N1 is in a Kozai-Lidov orbital resonance. S/2002 N5 was 25.9 mags, is
about 23 km in size and it makes a family of distant prograde
satellites with Sao and Laomedeia. This survey mostly completes the
outer satellites of Uranus to about 8 km and Neptune to about 14 km in
diameter. The size distributions of satellite dynamical families
around the giant planets shows a strong steepening in the power law
size distribution smaller than 5 km in diameter. The satellites of a
family become much more common smaller than 5 km and their size
distribution is consistent with a collisional break-up of a once
larger parent satellite.

\end{abstract}


\section{Introduction}

The giant planets are known to have many satellites that can be
classified as either small inner prograde satellites, medium to large
regular prograde satellites or small outer irregular satellites. The
regular satellites, like the four large Galilean satellites around
Jupiter, are thought to have formed with the planet in a
circumplanetary disk of gas and dust (Canup \& Ward 2002; Cuk et
al. 2020a; Batygin \& Morbidelli 2020). The small inner satellites are
close to the planet where collisions, tidal forces and the Roche
radius limit of the planet creates a chaotic environment that can
perturb and/or disrupt these satellites over the age of the solar
system (Cuk et al. 2020b, 2022; Kane \& Li 2023). Many of the small
inner giant planet satellites are associated with rings of the giant
planets (French et al. 2015; Charnoz et al. 2018). The regular
satellites of the planets are generally medium to large satellites
that can easily be observed with modern telescopes (Salmon \& Canup
2017; Neveu \& Rhoden 2019). The inner small satellites are best
discovered by spacecraft like Voyager 2 and Cassini or spaced-based
observatories like the Hubble Space Telescope, which have small
fields-of-view, but high resolving and deep imaging power to deal with
the strong scattered light near a planet (Showalter et al. 2019).

The outer satellites have irregular, distant, inclined and eccentric
orbits that can be either prograde or retrograde, which suggests they
did not form with the planet, but were captured during or just after
the planets formed (Jewitt \& Haghighipour 2007; Nicholson et
al. 2008; Nesvorny 2018). Current capture of a satellite by a planet
is not efficient in the solar system, but in the past several
mechanisms were operating more prominently that could remove energy
from the orbit of a passing object to make satellite capture more
probable.  These capture mechanisms include close planet-planet
encounters (Cuk \& Gladman 2006; Nesvorny et al. 2007), gas drag (Cuk
\& Burns 2004), and collisions or three body interactions within the
Hill Sphere of a planet of passing comets or asteroids (Colombo \&
Franklin 1971; Vokrouhlicky et al. 2008; Philpott et al. 2010; Koch \&
Hansen 2011; Gaspar et al. 2013).

The number and configuration of planetary outer satellites are
interesting in the context of planet formation and migration (Jewitt
\& Sheppard 2005; Jewitt \& Haghighipour 2007; Nicholson et al. 2008;
Nesvorny 2018).  Remarkably, all the giant planets appear to have
similar outer satellite systems for the largest few satellites ($> 20$
km), even though these planets had different formation histories
(Sheppard et al. 2006). This suggests the capture of these outer
planetary satellites occurred just after the planet formation epoch,
and their capture was independent of the planet's mass or formation
location. The outer satellites could be windows into the planet
building and migration process as they were likely captured during the
Solar System's earliest days.

Jupiter currently has 95 known satellites of which 87 are outer
irregular satellites, Saturn has 146 with 122 outer satellites,
Uranus, with this work, now has 28 with 10 outer satellites and
Neptune 16 with 8 outer satellites. Many of the outer satellites of
Jupiter and Saturn have been shown or thought to be in dynamical
orbital families, suggesting there were originally only a few outer
parent satellites that had broken apart from collisions with
asteroids, comets or other satellites over the age of the solar
system. Jupiter appears to have 7 or 8 dynamically unique outer
satellite type orbits (Sheppard \& Jewitt 2003; Beauge \& Nesvorny
2007; Brozovic \& Jacobson 2017; Sheppard et al. 2018, 2023), while
Saturn has 5 or more dynamically unique outer satellite type orbits
(Gladman et al. 2001; Holt et al. 2018; Ashton et al. 2022; Jacobson
et al. 2022; Sheppard et al. 2023). Until now, it was not clear if
Uranus and Neptune had dynamical outer satellite families since only
the largest have been found to date. It has been suggested the Uranian
satellites Caliban and Stephano could be the largest members of a
group, but having just two objects makes this suggestion indeterminate
(Nesvorny et al. (2003); Kavelaars et al. (2004); Sheppard et
al. (2005)).  Neptune also has no obvious groupings of its largest
outer satellites, though the retrogrades Neso and Psamathe have
similar orbits suggesting a grouping as mentioned in Sheppard et
al. (2005). The outer Neptune prograde satellites Sao and Laomedeia
could also make a possible grouping (Holman et al. 2004). If these are
true groups at Uranus and Neptune, it would be expected that many more
smaller satellites should exist with similar orbits, like seen for
Jupiter and Saturn.

\begin{figure}
\epsscale{0.4}
\centerline{\includegraphics[angle=180,totalheight=0.7\textheight]{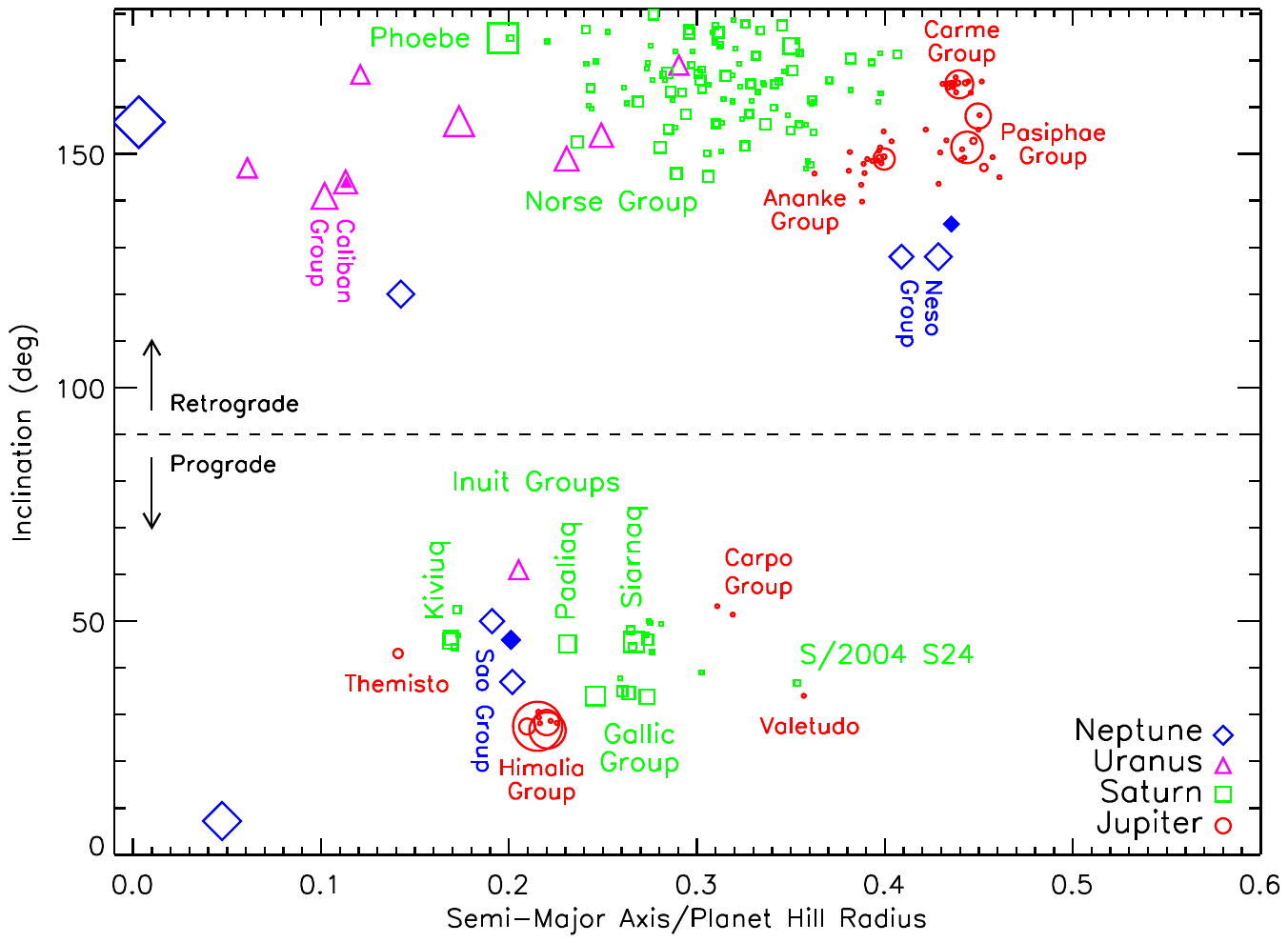}}
\caption{The known outer satellites of the giant planets where
  Neptunians are blue diamonds, Uranians magenta triangles, Saturnians
  green squares and Jovians red circles. Symbol size is proportional
  to the Log of the diameter of the satellite. Semi-major axis and
  inclination are the mean osculating orbital elements from this work
  for Uranus and Neptune, from Jacobson et al. (2022) for Saturn and
  Brozovic \& Jacobson (2017) for Jupiter with updates from new
  discoveries at Jupiter and Saturn reported in Sheppard et al. (2018,
  2023) and Ashton et al. (2020, 2021, 2022) shown at
  \href{https://ssd.jpl.nasa.gov/sats/elem}{ssd.jpl.nasa.gov/sats/elem}. The
  newly discovered Uranus and Neptune satellites reported here are
  shown as filled solid symbols while already known satellites are
  shown by open symbols. As seen here, all known outer satellites have
  semi-major axes less than 0.5 Hill radii. Dynamical families are
  identified by the largest member. The new Uranian satellite S/2023
  U1 has a similar orbit as Caliban and Stephano, making this the
  first group of three or more satellites known at Uranus.  The new
  Neptunian S/2021 N1 has a similar distant retrograde orbit as Neso
  and Psamathe, while the new S/2002 N5 has a similar prograde orbit
  as Sao and Laomedeia, making both these the first known groups of
  three or more satellites at Neptune.
  \label{fig:Allai2024JupSat}}
\end{figure}

Uranus and Neptune both have peculiarities that might have disrupted
their satellite systems. Uranus is tipped on its side with a $> 90$
deg obliquity, likely from a giant impact during the planet formation
process (Parisi et al. 2008; Rufu \& Canup 2022).  Uranus’ outer
satellites do appear somewhat unique in that the retrograde satellites
are closer to the planet and Uranus does not have a significant
prograde outer satellite population
(Figure~\ref{fig:Allai2024JupSat}).  Neptune likely captured the large
Kuiper Belt object Triton, which is bigger than Pluto. Triton, the
largest retrograde satellite by far in the solar system ($\sim 2700$
km), possesses a near circular and relatively close-in orbit to
Neptune.  Intriguingly, Nereid, the largest distant outer satellite of
Neptune at about 360 km, has the most eccentric orbit of any known
satellite, but interestingly a low inclination and relatively small
semi-major axis. This suggests Nereid might have once been an inner
satellite that formed with Neptune and was disrupted from Triton's
capture. In particular, the origin of Triton can unveil important
constraints on the formation or survival of Nereid and the remaining
smaller outer satellites (Agnor \& Hamilton, 2006; Rufu \& Canup
2017). Triton and/or Nereid could be captured objects from the Kuiper
Belt or ejected former inner satellites (Nogueira et al. 2011; Li \&
Christou 2020; Gomes \& Morbidelli 2024). The remaining outer
satellites of Neptune were probably captured from heliocentric orbit
as they have more traditional irregular outer satellite type orbits
(Holman et al. 2004; Sheppard et al. 2006; Nesvorny et al. 2014; Li et
al. 2020).

The size distribution of Uranus and Neptune satellites are poorly
understood because only the brightest and thus largest are known. For
this reason, models of collisional evolution of giant planets’ outer
satellites do not usually considered Neptune while collisional models
use an incomplete Uranus satellite system (Bottke et al. 2010).
Shallow size distributions are seen for the outer satellites between
20 and 100 km for Jupiter, Saturn, Uranus and Neptune (Sheppard et
al. 2006).  There is a steepening in the distribution at sizes less
than 15 km at Jupiter and Saturn (Sheppard \& Jewitt 2003; Nicholson
et al. 2008; Sheppard et al. 2018, 2023; Ashton et al. 2020,
2021). This is a sign of strong collisional evolution at Jupiter and
Saturn as groups of these numerous smaller satellites tend to have
dynamically similar orbits, showing a fragmentation of a once larger
satellite (Figure~\ref{fig:Allai2024JupSat}). Bottke et al. (2010)
required substantial depletion of captured outer satellites via
collisional evolution to satisfy observations.

For the much more numerous small outer satellites of only a few km in
size, it is unknown if Uranus and Neptune continue the trend of
similar outer satellite systems as Jupiter and Saturn since it is very
hard to discover small satellites at Uranus' and Neptune's extreme
distances. Jupiter's outer satellites are complete to about 2 km and
Saturn to about 3 km because of their closer distance to Earth
(Sheppard et al. 2023).  To better compare Uranus' and Neptune's
system to the other planets, we must search for satellites smaller
then the power law size break that likely occurs below about 15 km in
satellite size. Discovery of smaller outer satellites will help to
better understand the collisional evolution in these systems and
better compare them to the similar Trojan asteroids and Kuiper Belt
(Bottke et al. 2023a). In addition, future spacecraft missions to
Uranus or Neptune may be able to image some of the outer satellites of
these planets (Denk \& Mottola 2019; Cartwright et al. 2021; Cohen et
al. 2022).

For the outer satellites of the giant planets, the sky area needed to
search for stable satellites around a planet is so large, they can
only efficiently be discovered by large field-of-view ground-based
telescopes. The last successful ground-based surveys for satellites of
Uranus and Neptune were done about two decades ago, reaching about
25.5 to 26th magnitude in the r-band and able to detect satellites
larger than about 15 and 30 km around Uranus and Neptune, respectively
(Kavelaars et al. 2004; Holman et al. 2004; Sheppard et al. 2005,
2006; Brozovic et al. 2011). We here report on a new survey of the
region around Uranus and Neptune to extra ultra-deep depths using the
Magellan 6.5m and Subaru 8.2m telescopes, going more than a magnitude
fainter than previous observations to over 27th mags. This very
ultra-deep survey was done by imaging the Hill spheres of Uranus and
Neptune with tens of images over several hours on several different
nights and then shifting and adding the images together at the
planets' motion to detect faint satellites not normally visible in a
single image.

\section{Observations and Methods}

We used the Subaru and Magellan telescopes to observe most of the
dynamically stable Hill Spheres of Uranus and Neptune to extra
ultra-faint depths through a pencil-beam survey.  Subaru has a 8.2
meter primary mirror and uses the Hyper Suprime-Cam (HSC) imager at
prime focus (Miyazaki et al. 2018; Komiyama et al. 2018; Kawanomoto et
al. 2018; Furusawa et al 2018). HSC covers about a 1.5 degree diameter
and has 112 individual CCDs that are $2048 \times 4096$ pixels. The
pixel size is 0.168 arcseconds per pixel. The r-band filter was used.
We also used the 6.5 meter Magellan telescope for discovery and
recovery with the Inamori Magellan Areal Camera and Spectrograph
(IMACS) camera at the F/2 focus, which has 8 CCDs of $2048 \times
4096$ pixels and a pixel scale of 0.20 arcseconds per pixel (Dressler
et al. 2011). The very wide band filter WB4800-7800 was used, which
covers a wavelength range from 4800 to 7800 angstroms that is similar
to a very broad VR filter. The IMACS field-of-view is about 0.17
square degrees.

Neptune's and Uranus' Hill Spheres, where satellites are theoretically
stable, are about 1.5 degs (0.78 au) and 1.4 degs (0.49 au) in radius
on the sky, respectively. But simulations, analytical analysis and
empirical data from other planets show that outer satellites are only
stable to about 0.7 Hill radii for retrograde and 0.5 Hill radii for
prograde satellites, making the true area that satellites are stable
at about 1 degree radius or less on the sky (Hamilton \& Krivov 1997;
Sheppard et al. 2006; Shen \& Tremaine 2008; Donnison 2011). In
addition, projection effects of a satellite orbit onto the sky makes
even the largest satellite orbits usually appear well less than 0.7
Hill radii away from the planet on sky, even when the satellites are
near the extremes of their orbits. HSC covers most of the stable
Uranus or Neptune Hill Spheres in one image when the planet is placed
near the center of the field, but IMACS does not and thus requires
multiple pointings to cover the primary inner Hill Sphere of a giant
planet.

The extra ultra-faint magnitudes we need to achieve are not possible
in a single exposure due to the background saturation of the CCDs that
would occur after several minutes of exposing. The satellites also
move relative to the background stars and galaxies, making the
satellites trail after just a few minutes of exposure when guiding the
telescope at sidereal rates. Because of the background saturation
problem in very long exposures and the movement of the satellites at
non-sidereal rates, we used a shifting and adding/medianing of the
images taken during a single night to get to extra ultra-faint
magnitudes that we required to explore the smaller sized satellite
populations around these planets. The Image Reduction and Analysis
Facility (IRAF) with the imshift and imcombine routines was used on
images from a given night to shift them by the apparent motion of the
respective planet and then add them together to get one single extra
ultra-deep image. The same images were also shifted and medianed
together to get a second type of extra ultra-deep image. In such a
deep image the stars and galaxies appear trailed, while any object
with a motion similar to the planet will be seen as a point source,
making very faint satellites discernible from the background
noise. Using this time intensive technique on some of the largest
telescopes in the world allowed the images to reach fainter magnitudes
around Uranus and Neptune then any previous observations.

\begin{figure}
\epsscale{0.4}
\centerline{\includegraphics[angle=0,totalheight=0.7\textheight]{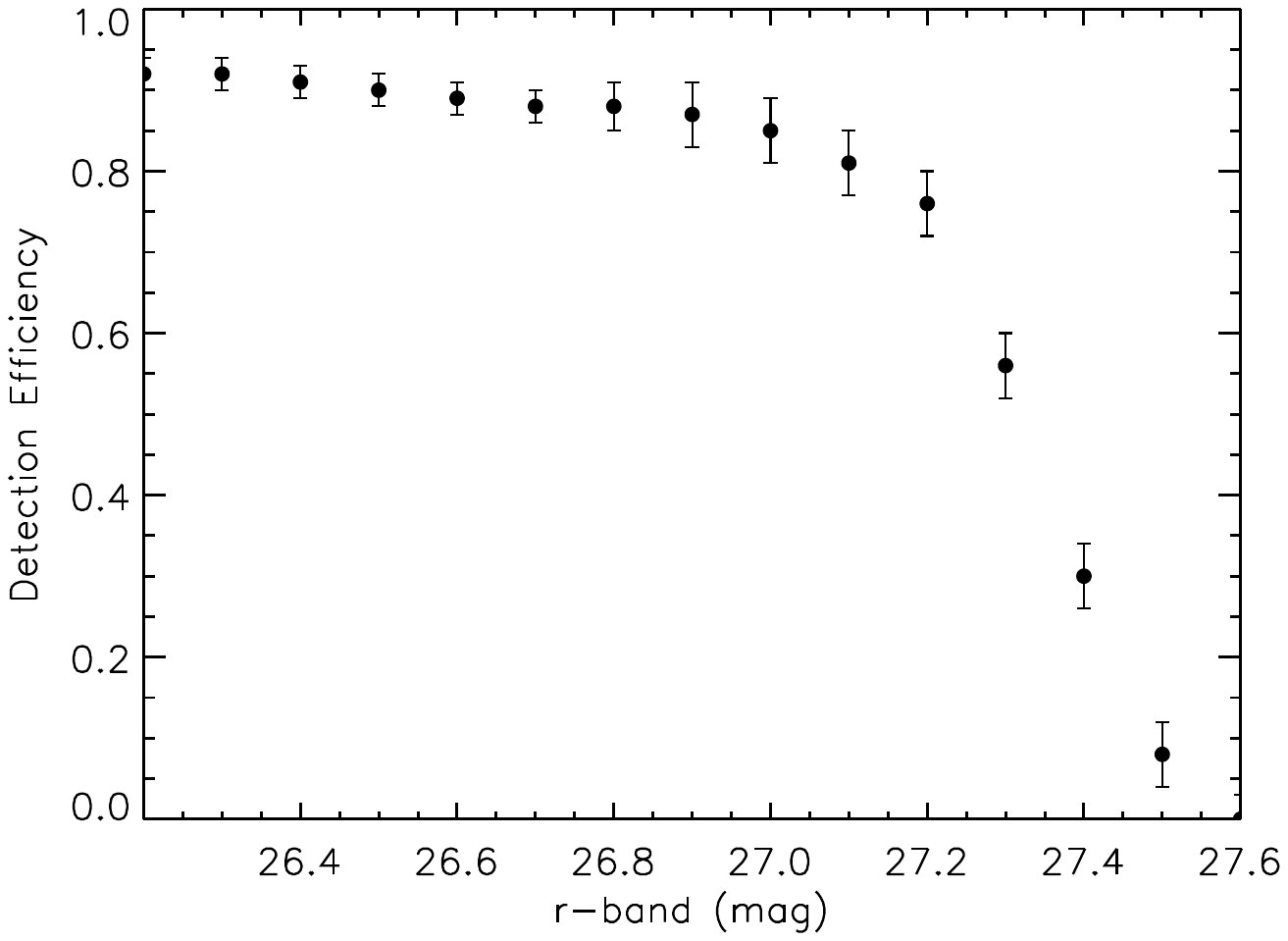}}
\caption{The efficiency of searching the Subaru September 7, 2021 data
  for Neptune satellites. We would expect to find over 75 percent of
  satellites within the field of view of HSC at 27.2 magnitudes in the
  r-band, which we take as the limiting magnitude of this one
  night. Table 1 shows all the detection efficiencies for the various
  nights of observations at Uranus and Neptune.
  \label{fig:efficiency2021nept}}
\end{figure}

Satellites were searched for in three complementary ways in all of the
data taken. First, the co-added images were examined by eye for any
point sources.  Second, the median images were examined by eye for any
point sources. The co-added and medianed images were then compared at
the location of any detected point sources to determine if the source
was likely to be real or just a hot pixel or cosmic ray, which appear
more prominent in the the co-added images compared to the medianed
images. Finally, the first half and second half of each night's data
were co-added as well as medianed together and then visually compared
to look for moving objects, which the eye can identify at
Signal-to-Noise levels of about 3.5. All these techniques have similar
depths of identifying real objects, with the medianed images usually
giving slightly better results. We used the known Neptune and Uranus
satellites found in the data along with implanting fake objects into
the data to determine our limiting magnitude for each observation. As
done in previous moving object surveys, we used the imexam routine in
IRAF to determine the point-spread function of each image to generate
artificial objects with the IRAF routine mkobjects that were implanted
into the images (Sheppard and Trujillo 2016).
Figure~\ref{fig:efficiency2021nept} shows the typical efficiency curve
of finding moving objects in a median single night dataset, which
gives us our limiting magnitudes for each nights data shown in
Table~\ref{tab:fields2024}.

With deep recovery imaging of newly found satellites, we duplicated
large areas of coverage around both Uranus and Neptune in three
different years, 2021, 2022 and 2023 at multiple different times
during a single year. This means satellites that may have been
originally obscured by a background star or out of the field-of-view
in one month or year would likely not be so in a following time,
making the survey more complete. This completeness is demonstrated by
our recovery multiple times of all known outer satellites of both
Uranus and Neptune, without prior knowledge of their locations.

\subsection{Neptune Satellite Pencil Beam Survey}
\begin{figure}
\epsscale{0.4}
\centerline{\includegraphics[angle=0,totalheight=0.7\textheight]{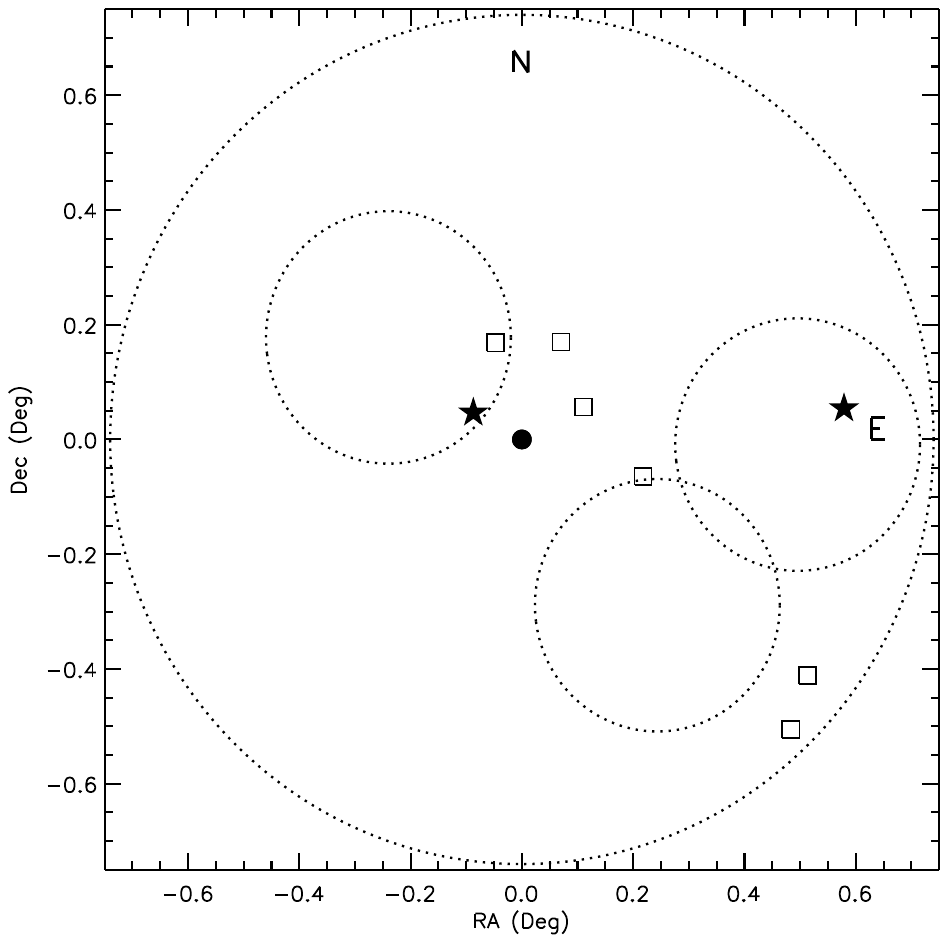}}
\caption{Area searched around Neptune in 2021. Known outer Neptune
  satellite positions are shown as open square symbols for September
  7, 2021. Neptune is at the center of the field shown by the filled
  circle. The large dotted circle is the field of view of
  HyperSuprime-Cam on Subaru. The smaller dotted circles are the
  fields imaged with IMACS on Magellan that went to a depth beyond
  26.5 mags taken on September 3 and October 6 in 2021 (see
  Table~\ref{tab:fields2024}). All known outer satellites of Neptune
  were detected. The newly discovered Neptune satellites S/2021 N1
  (near far East edge) and S/2002 N5 (near Neptune) are shown by
  filled stars.
  \label{fig:areaneptune2021}}
\end{figure}

Subaru HSC was used on the nights of UT September 7 and 8, 2021 with
Neptune placed near the center of the field-of-view. Nineteen images
of 350 seconds were taken on September 7 and twelve images on
September 8 (Figure~\ref{fig:areaneptune2021}). The average seeing for
the Neptune images on September 7th was 0.83 arcseconds and 0.77
arcseconds on September 8th (see Table~\ref{tab:fields2024}). S/2021
N1 was found on both nights of the Subaru Neptune observations
(Sheppard et al. 2024a), as was S/2002 N5 on both nights, which was
already found using Magellan a few days prior (Sheppard et al. 2024b).

We used Magellan to image to the West and East sides of Neptune on UT
3 September 2021, putting Neptune just outside the field-of-view for
both long stares. The West imaging obtained sixteen images of 380
seconds in 0.73 arcsecond seeing and the East imaging had twenty-one
images in 0.67 arcsecond seeing (Table~\ref{tab:fields2024}). S/2002
N5 was found in the Neptune West images after co-adding them (Sheppard
et al. 2024b).

After the main discovery images around Neptune from Subaru and
Magellan in September 2021, we used Magellan again in October and
December 2021 to recover the newly found Neptune satellites (Table
~\ref{tab:fields2024}). During the recoveries, we again searched the
complete IMACS field-of-view for any possible missed satellites during
the discovery images in September 2021, with nothing new found. The
two new Neptune satellites were again imaged in 2022, the brighter
S/2002 N5 at Magellan and the fainter S/2021 N1 at the VLT using the
FORS2 imager. Gemini with the Gemini Multi-Object Spectrograph (GMOS)
imager was used on UT 3 November 2023 to confirm the faintest of the
new Neptune satellites (Hook et al. 2004). Both new Neptune satellites
were imaged at Magellan on UT 4 November 2023 to fully determine their
orbits and they were announced on Minor Planet Electronic Circulars in
February 2024 (Sheppard et al. 2024a, 2024b). All recovery images were
fully searched to look for additional new satellites, with no new
satellites found.

As shown in Figure~\ref{fig:efficiency2021nept} the Neptune Subaru
survey on September 7 had a limiting magnitude of 27.2 mags while the
September 8 observations, which repeated the same area of space around
Neptune, had a limiting magnitude of 26.9 mags. In addition, several
Magellan ultra-deep images were taken near Neptune with depths as
faint as 26.9 mags (see Table~\ref{tab:fields2024}).  All known
Neptune outer satellites were detected on multiple nights in these
observations and astrometry reported to the Minor Planet Center shown
in Table~\ref{tab:residsnep} as many had not been observed for several
years, some not since 2009. From these observations, the outer
satellites of Neptune should be nearly complete to about 27 mags,
corresponding to satellites of about 14 km in diameter assuming
albedos of 0.1.

\subsection{Uranus Satellite Pencil Beam Survey}

Uranus was placed near the center of the field-of-view using Subaru on
8 September 2021 with twelve images of 300 seconds taken in average
seeing of 0.73 arcseconds (see Table~\ref{tab:fields2024} for
details). The Uranus images using Subaru were not as deep as those for
Neptune since only twelve images of 300 seconds were taken on only one
night for Uranus, while Neptune had nineteen images of 350 seconds on
one night and another twelve on a second night. In addition, Uranus
was off-opposition in September, which amounted to losing about 0.3
magnitudes in brightness for the satellites from the increased
distance from Earth and the higher phase angle.  In the co-added and
medianed images, a very faint object was identified as a point source
and moving at Uranus' rate, but was too faint to be a reliable
detection to use significant large telescope time to follow-up. A few
brighter candidates that were moving a little off of Uranus' motion
were imaged at Gemini a month later and found to not be Uranus
satellites.

\begin{figure}
\epsscale{0.4}
\centerline{\includegraphics[angle=0,totalheight=0.7\textheight]{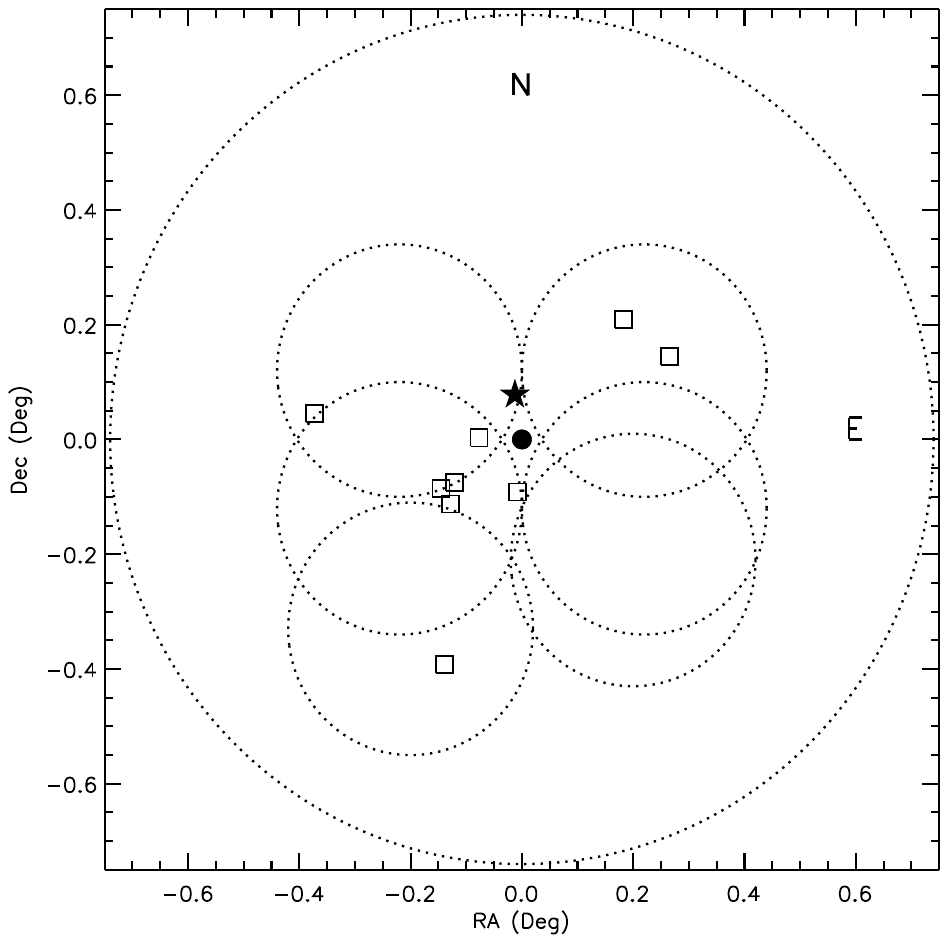}}
\caption{Area searched around Uranus in 2021. Known outer Uranus
  satellite positions are shown as open square symbols for December 2,
  2021. Uranus is at the center of the field shown by the filled
  circle. The large dotted circle is the field of view of
  HyperSuprime-Cam on Subaru and the smaller dotted circles are the
  fields imaged with IMACS on Magellan (see
  Table~\ref{tab:fields2024}). The Southeast quadrant has two, mostly
  overlapping, fields since this area was re-imaged in December 2021
  with IMACS as the October 2021 observations had poor seeing. All
  known outer satellites of Uranus were detected. The newly discovered
  Uranus satellite S/2023 U1 is shown by a filled star.
  \label{fig:areauranus2021}}
\end{figure}

\begin{figure}
\epsscale{0.4}
\centerline{\includegraphics[angle=0,totalheight=0.7\textheight]{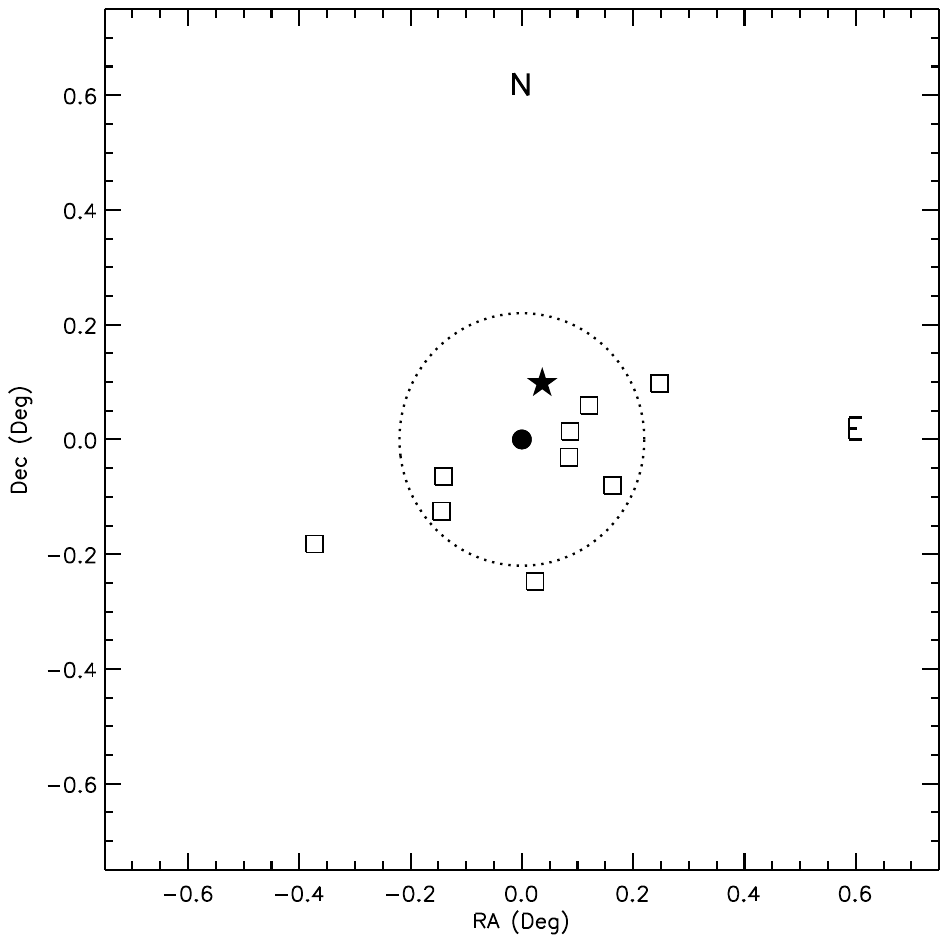}}
\caption{Area searched around Uranus in 2023. Known outer Uranus
  satellite positions are shown as open square symbols for November 4,
  2023. The dotted circle is the field-of-view imaged with IMACS on
  Magellan on three different nights in November and December 2023
  with Uranus placed at the center of the field (see
  Table~\ref{tab:fields2024}). The newly discovered Uranus satellite
  S/2023 U1 is shown by a filled star. The Magellan observations of
  2023 had 6 of the 9 known outer satellites of Uranus easily detected
  with only the most distant satellites outside the field of view
  undetected.
  \label{fig:areauranus2023}}
\end{figure}

Fields around Uranus were further imaged using Magellan to extra
ultra-faint depths in late October and early December 2021, allowing
all four quadrants around Uranus to be again deeply imaged using
Magellan (Figure~\ref{fig:areauranus2021}). The Magellan imaging cycle
for one field lasted 2 to 4 hours on each night
(Table~\ref{tab:fields2024}). No obvious new satellites of Uranus were
detected in the 2021 images, though a faint almost in the noise source
was flagged in the Magellan images, but it was near a bright star and
deemed too faint and questionable as a source to use a significant
amount of large telescope time to follow-up. Since there might be very
faint unconfirmed satellites around Uranus, the main space around
Uranus was again imaged for new satellites at Magellan on 4 November
2023 with Uranus placed at the center of the field-of-view
(Figure~\ref{fig:areauranus2023}). One new candidate Uranus satellite
was detected and follow-up imaging at Magellan on December 6 and 13
showed this source as likely being a new Uranian satellite. A basic
Uranian satellite orbit was determined for the newly found object
based on the 3 observations in late 2023, which was able to link the
very faint sources noticed at both Subaru and Magellan in the 2021
data to the 2023 observations. With over two years of observations,
S/2023 U1 was announced on a Minor Planet Electronic Circular in
February 2024 (Sheppard et al. 2024c).

All known Uranus outer satellites were easily detected in our survey
images on at least two different observing nights without prior
knowledge of their locations. The new astrometry shown in
Table~\ref{tab:residsura} was reported to the Minor Planet Center as
some of these known satellites had not been observed since 2004. We
covered the inner 77 percent of Uranus' stable Hill sphere radius to
depths of 26.5 to 26.9 mags. The very inner region around Uranus where
most satellites would expect to be found was observed with overlapping
fields in 2021 as well as having our deepest Magellan image centered
on Uranus in 2023 to 26.9 mags.  The outer satellites of Uranus should
be nearly complete to about 26.5 mags, corresponding to satellites of
about 8 km in diameter assuming albedos of 0.1.

\section{Results}

We found three new satellites, one around Uranus and two around
Neptune. The new Uranus and Neptune satellites are the faintest ever
discovered around the planets using ground-based telescopes. In
addition, we also detected all known outer satellites of Uranus and
Neptune on multiple nights, some not seen since 2004.

All of the new Uranian and Neptunian satellites have distant,
eccentric and inclined orbits that suggests they were captured
satellites, which likely occurred during or just after the planet
formation epoch (Figure~\ref{fig:Allai2024JupSat}).  These new
observations around the outer planets nearly complete the satellite
inventories of Uranus and Neptune to about 26.5 and 27 mags or about 8
and 14 km, respectively, assuming albedos of the satellites of ten
percent (see Table~\ref{tab:fields2024}). Jupiter is complete to about
2 km in size, while Saturn is complete to about 3 km in size for
satellites, with the closer planets better positioned to find fainter
and thus smaller satellites (Sheppard et al. 2023).

\subsection{New Uranian Satellite S/2023 U1}

The newly discovered Uranus satellite is provisionally named S/2023
U1, and this now gives Uranus 28 known satellites, of which 10 are
outer irregular satellites.  The new Uranian satellite has over 2
years of observations and thus a secure well-determined orbit around
Uranus (Table~\ref{tab:newmoons}).  S/2023 U1 is about 7 km in
diameter assuming an albedo of ten percent, likely making it the
smallest satellite ever observed around Uranus. It has a distant and
inclined retrograde orbit around Uranus, taking about 1.86 years to
orbit Uranus once. This orbit is similar to Uranian outer satellites
Caliban and Stephano. S/2023 U1 will receive a permanent number and
name based on a Shakespeare character.

\subsection{New Neptunian Satellites S/2021 N1 and S/2002 N5}

Two new satellites of Neptune were discovered. Neptune now has 16
known satellites, of which 8 are outer irregular satellites plus
Triton. The fainter Neptune satellite has a provisional designation
S/2021 N1 and is the faintest satellite ever discovered at 26.9 mags
by ground-based observations. S/2021 N1 is about 14 km in diameter
with a distant retrograde orbit around Neptune of almost 27 years
(Table~\ref{tab:newmoons}). S/2021 N1 has an orbit that is
similar to Neptunian outer satellites Neso and Psamathe.

Once an orbit around Neptune was determined for the brighter new
Neptune satellite using the 2021, 2022 and 2023 observations, it was
traced back to an object that was spotted near Neptune in 2002, but
lost before it could be confirmed as orbiting Neptune (Holman et
al. 2004). The brighter Neptune satellite now has a provisional
designation S/2002 N5 with a prograde orbit of almost 9 years to orbit
Neptune and is about 23 km in diameter
(Table~\ref{tab:newmoons}). S/2002 N5 has an orbit that is similar to
Neptunian outer satellites Sao and Laomedeia.  The new Neptune
satellites have over two years of observations and thus secure well
determined orbits and will get permanent numbers and names based on
the fifty Nereid sea goddesses in Greek mythology.

\subsection{Keplerian Osculating Orbital Elements of Outer Satellites}

The outer satellites of the giant planets have significant
gravitational interactions with the other giant planets as well as the
Sun. Thus the orbits of the satellites are not closed orbits but may
vary over time.  Using the new 2021, 2022 and 2023 astrometry obtained
on all of the outer satellites of Uranus and Neptune, we performed
numerical integrations to determine the average, minimum and maximum
variations (osculating elements) in the orbits of the known outer
satellites over 10,000 years. Results are the Neptune satellite orbits
nep104 and Uranus satellite orbits ura117 in the JPL {\it Horizons}
On-Line Solar System Data Service (Giorgini et al. 1996) and from
NASAs Navigation and Ancillary Information Facility (Acton 1996)
(\href{https://ssd.jpl.nasa.gov/sats}{ssd.jpl.nasa.gov/sats/ephem/files.html}).
For the numerical simulations, we use the same parameters and
techniques as detailed in Brozovic \& Jacobson (2022). This dynamical
model was also previously used for numerically integrated ephemerides
of the outer irregular satellites in Jacobson et al. (2012), Brozovic
\& Jacobson (2017) and Jacobson et al. (2022). In brief, we add the
masses of Mercury, Venus, the Earth-Moon system and Mars as part of
the Sun's mass. For Uranus additional perturbers used are the planet
Uranus, Uranus J2, Uranus J4 and its largest satellites Miranda,
Umbriel, Ariel, Oberon and Titania along with the Jovian system,
Saturnian system, and Neptunian system. For Neptune additional
perturbers used are the planet Neptune, Neptune J2, Neptune J4,
Neptune J6 and its largest satellite Triton as well as the Jovian
system, Saturnian system and Uranian system (see Brozovic \& Jacobson
(2022) for full information). The JPL planetary ephemeris DE441 is
used to determine the location of the planets and Sun and simulate the
long-term dynamics of the satellites (Park et al. 2021).

\begin{figure}
\plotone{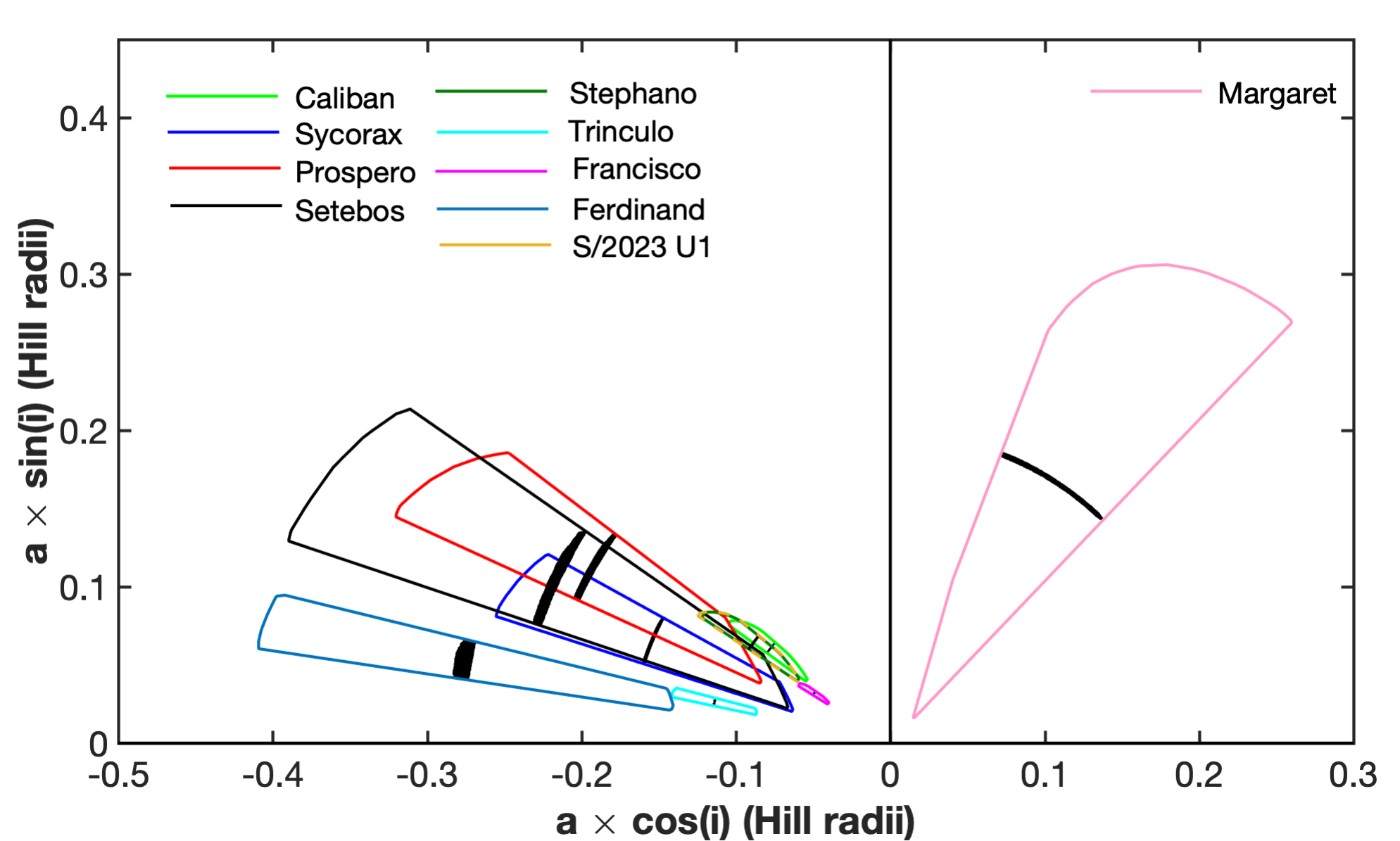}
\caption{Orbital phase space for the irregular satellites of Uranus.
  Black dots show osculating semi-major axis multiplied by sines and
  cosines of the osculating inclination, while the colored outlines
  show the pericenter--to--apocenter variation, $2ae$, for the
  duration of 10000 years. The reference plane for inclinations is the
  ecliptic. The Hill radius for Uranus is 0.47 au. The newly
  discovered satellite S/2023 U1 (in orange) completely overlaps in phase space
  with Stephano.
  \label{fig:uranus_hill}}
\end{figure}

\begin{figure}
\plotone{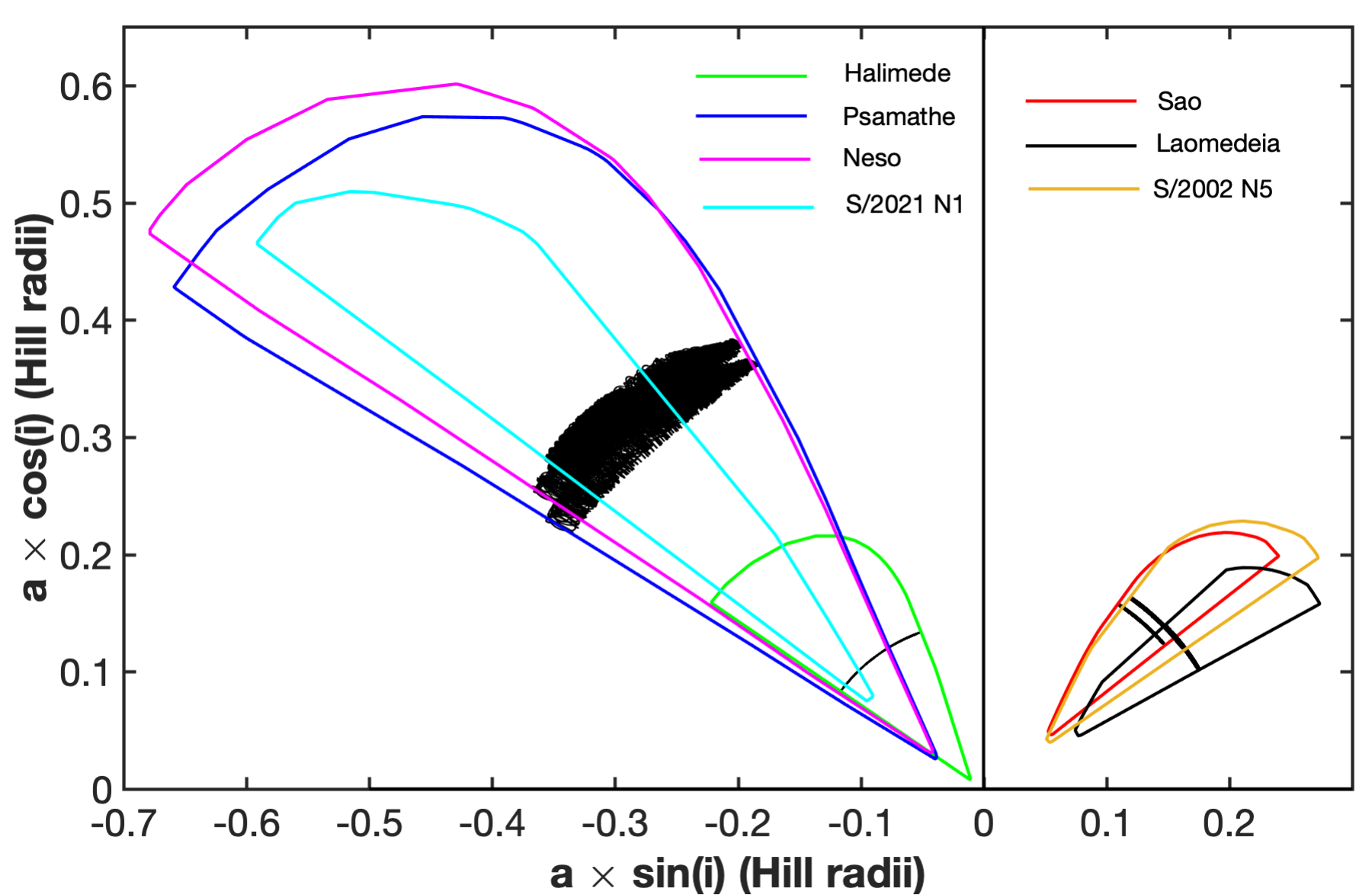}
\caption{Orbital phase space for the irregular satellites of
  Neptune. Black dots show osculating semi-major axis multiplied by sines and
  cosines of the osculating inclination, while the colored outlines
  show the pericenter--to--apocenter variation, $2ae$, for the
  duration of 10,000 years. The reference plane for inclinations is the
  ecliptic. The Hill radius for Neptune is 0.77 au.
\label{fig:neptune_hill}}
\end{figure}

The elemental space of all of the outer satellites of Uranus and
Neptune based on their osculating orbits can be seen in
Figures~\ref{fig:uranus_hill} and~\ref{fig:neptune_hill}. A dynamical
grouping of satellites suggesting a common origin from a once bigger
parent body is likely if both the semi-major axis and inclination
overlap for two or more objects in their osculating orbital elements
over time. Tables~\ref{tab:uranus_osc} and~\ref{tab:neptune_osc} show
mean osculating elements for 10,000 years of orbit integration.

The osculating orbital elements show there are dynamical orbital
groupings of outer satellites around both Uranus and Neptune, like
seen at Jupiter and Saturn.  We find each of the newly discovered
satellites reported here are likely in a dynamical grouping (see
below), which show for the first time Uranian and Neptunian dynamical
groups containing each three members. This suggests once larger
parents satellites at Uranus and Neptune had broken apart due to past
collisions, most likely with other moons, comets or asteroids, leaving
the broken fragments behind in similar orbits as the original larger
satellite.  If these are really collisional remnants of once larger
satellites, it is likely many smaller satellites exist in these newly
identified Uranian and Neptunian groupings, like found at Jupiter and
Saturn, but they would be only a few km in size or smaller and too
faint to efficiently detect with current telescope and detector
technology.

\subsection{Dispersion Velocities of Satellite Dynamical Groupings}

Another way to determine dynamical groupings of objects is to examine
their dispersion velocities of their orbits relative to each
other. The initial dispersion velocity would be expected to be near
the escape speed of a disrupted body (Durda 2007). Most asteroid
families would be expected to have initial dispersion velocities less
than 100 m/s (Michel et al. 2015; Nesvorny et al. 2015). These initial
dispersion velocities would likely grow in time through various
perturbations over time (Carruba \& Nesvorny 2016; Li \& Christou
2018).

Collisional disruption of a small satellite around a planet is less
well understood than that for asteroids in heliocentric orbit
(Nesvorny et al. 2003; Bottke et al. 2024b). Not only are the physical
characteristics of the satellites poorly understood, but a satellite
around a planet may experience significantly different forces that
could increase the dispersion velocity between family members
(Nesvorny et al. 2004). Outer satellites of the planets experience
significant three-body interactions involving the Sun or other planets
as well as various resonances such as the Kozai-Lidov resonance that
we find to be operating on some, but not all, members of a dynamical
family (Carruba et al. 2002, 2004; Beauge \& Nesvorny 2007; Frouard et
al. 2011; Brozovic and Jacobson 2022).

The prograde Himalia family of outer satellites at Jupiter has a very
high dispersion velocity of up to 400 m/s (Li \& Christou 2017), while
the retrograde groups Carme and Ananke at Jupiter have about 50 and 80
m/s (Nesvorny et al. 2004). We assume any two objects with a minimum
dispersion velocity of less than about 100 m/s have a good chance of
being fragments from the same parent body.

We conducted a velocity dispersion simulation for the new Uranian and
Neptunian dynamical groupings. The following metric for distance
between elements of two different satellite was used (Beauge \&
Nesvorny 2007):
\begin{equation}
d^2_4=C_a (\frac{\Delta a}{a})^2+C_e(\Delta e)^2+C_I(\Delta sin(I))^2
\end{equation}

Where $\Delta a$ is difference in osculating semi-major axis between two satellites, $\Delta e$ is for eccentricity, and $\Delta sin(I)$ is for sinus of inclination. In addition, $C_a=\frac{(1-\bar{e}^2)^2}{4\bar{e}^2}$, $C_e=1/2$, and $C_I=2$.

The dispersion velocity for the group is defined as:
\begin{equation}
    (\delta V)^2=\frac{\mu}{\bar{a}(1-\bar{e}^2)}d_4^2
\end{equation}

where $\mu=GM_p$, $G$ is the gravitational constant, $M_p$ is
the mass of the planet, $\bar{e}$ and $\bar{a}$ are the averaged
eccentricity and averaged semi-major axis of all group members. We
used 30,000 years of osculating elements and calculated the dispersion
velocity every 100 days. Table~\ref{tab:dispvelocity} shows the
results of the dispersion velocity simulations of all likely members.
It is clear the three newest discoveries all have the lowest
dispersion velocities among group members, indicating that these
smaller satellites are probably collisional fragments of once larger
satellites.

\subsection{Uranian Caliban Dynamical Group}

\begin{figure}
\plotone{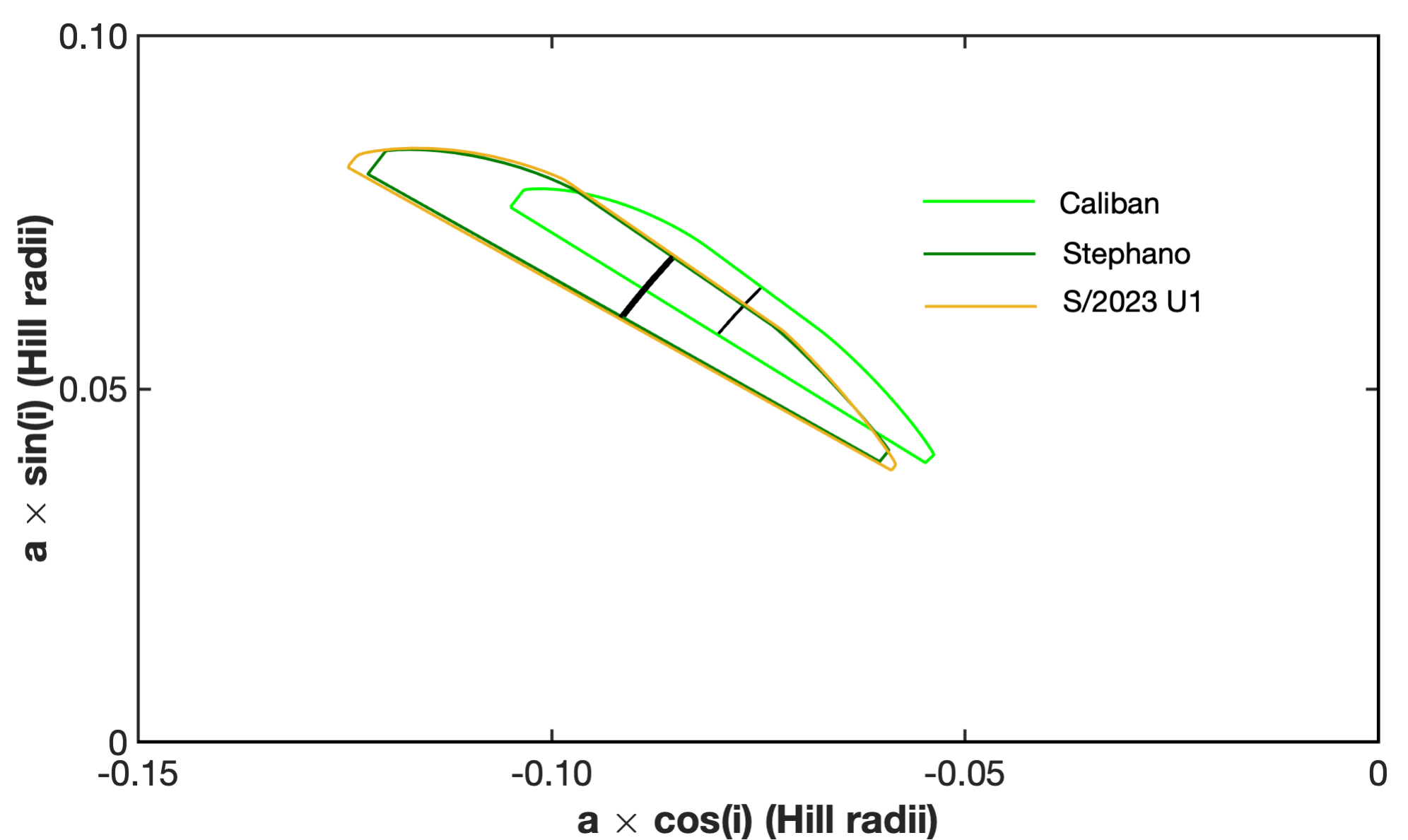}
\caption{Caliban group of satellites. The newly discovered satellite
  S/2023 U1 (in orange) completely overlaps in phase space with
  Stephano. We investigated how close they can approach to each other
  within 30,000 years of orbit integration: Stephano and S/2023 U1
  could have about a 22,000 km encounter around 5176 AD.
  \label{fig:calliban_group}}
\end{figure}

At Uranus, S/2023 U1 has a similar orbit as Caliban and Stephano
(Table~\ref{tab:uranus_osc}), making this a group of three satellites
that is labeled as the Caliban group in
Figure~\ref{fig:Allai2024JupSat}. In fact, S/2023 U1 has such a
similar orbit as Stephano they almost completely overlap the phase
space of each other (Figure~\ref{fig:calliban_group}).  Their minimum
dispersion velocity is only 21 m/s (Table~\ref{tab:dispvelocity}). We
find these objects come within about 22,000 km of each other in a few
thousand years. This is strong evidence that S/2023 U1 and Stephano
originated from the same parent body. Caliban, the other possible
group member, overlaps with S/2023 U1 and Stephano in inclination, but
has a slightly smaller semi-major axis that does not quite overlap
with the osculating orbital elements of Stephano and S/2023 U1. This
might suggest the smaller satellite S/2023 U1 was created from an
impact into Stephano after the original event that created Caliban and
Stephano. The Caliban family with S/2023 U1 shows that the outer
irregular satellites of Uranus likely have broken apart and it is
possible the dust produced from these events may create the red
material seen on the leading hemispheres of some of the inner large
Uranian satellites Titania, Oberon and Umbriel (Tamayo et al. 2013;
Cartwright et al. 2018, 2023; Graykowski \& Jewitt 2018).

Caliban has recently been observed in the infrared with a tentative
detection in the thermal from Herschel observations, measuring a
possible diameter of $42^{+20}_{-12}$ km (Farkas-Takacs et al. 2017;
Sharkey et al. 2023). Stephano is about half the size of Caliban and
S/2023 U1 much smaller at about 7 km in diameter. This would make the
effective diameter of the parent satellite that may have created this
dynamical group about 45 km in diameter, with Caliban having about $90
\%$ of the total volume of the three known members.

\subsection{Neptunian Neso Dynamical Group}
\begin{figure}
\plotone{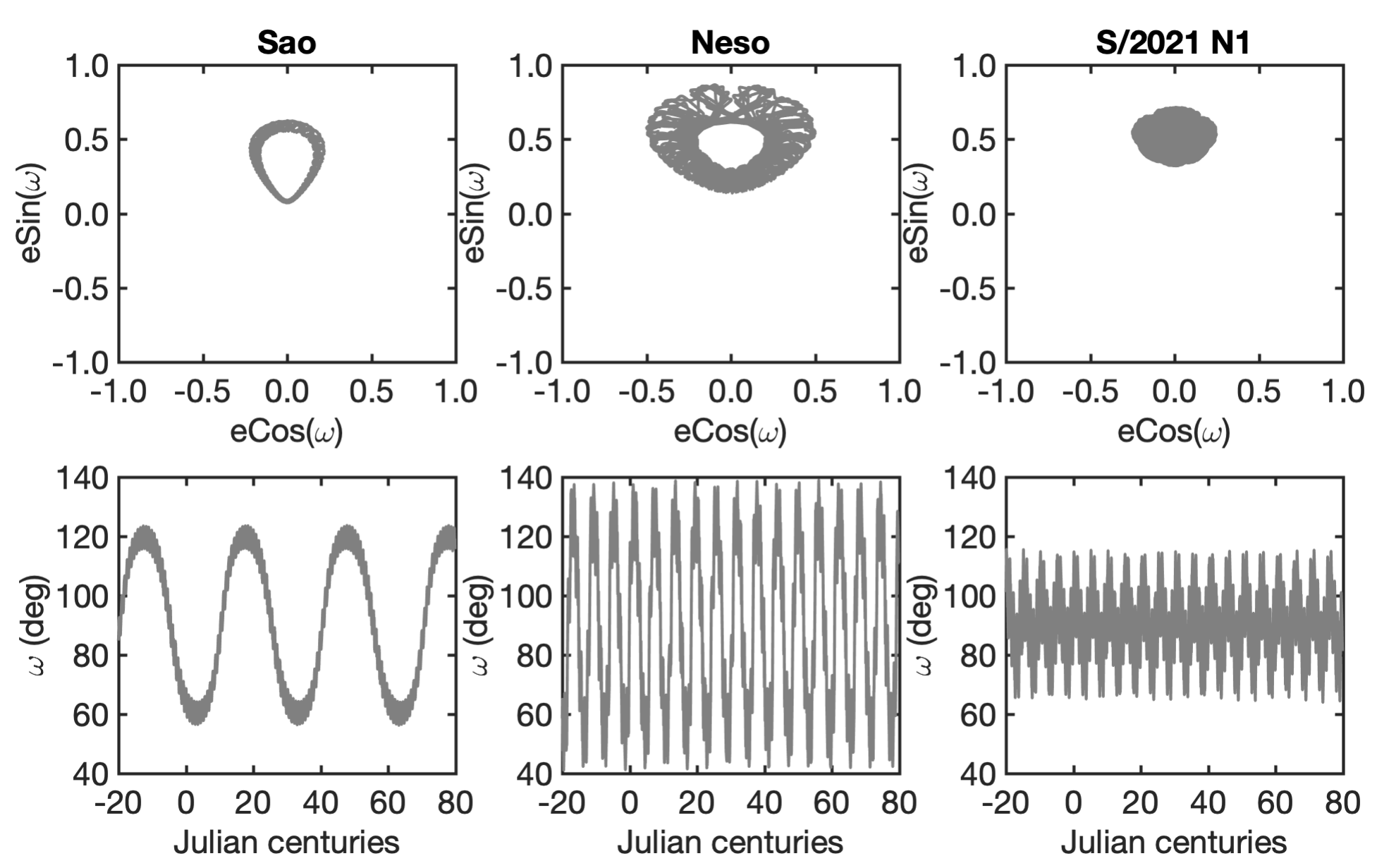}
\caption{Representation of the Kozai–Lidov dynamics for Neptune
  satellites Sao, Neso, and new S/2021 N1. The top panels show
  osculating ($e cos(\omega)$,$e sin(\omega)$), while the bottom
  panels show the osculating argument of pericenter ($\omega$) for
  10,000 years of orbit integration.  The argument of pericenter
  librates around 90 deg. The reference plane for inclination is the
  orbital plane of Neptune around the Sun with the mean pole of RA
  273.46 and Dec 67.71 degrees. The mean pole is estimated from de441
  planetary ephemeris.
  \label{fig:kozai2}}
\end{figure}

\begin{figure}
\plotone{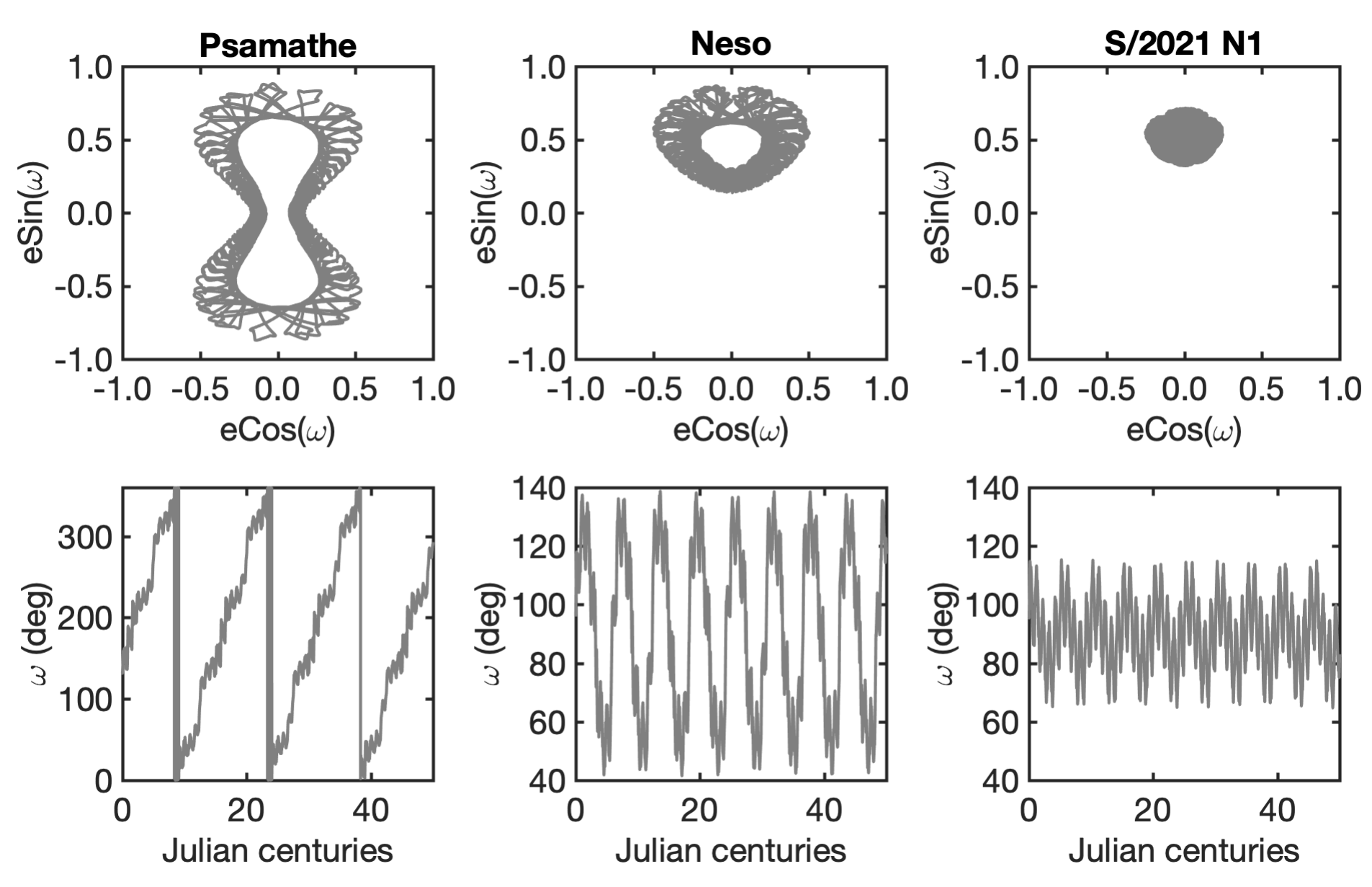}
\caption{Same as Figure~\ref{fig:kozai2} but for the Neso group of
  Neptune satellites. Neso and new satellite S/2021 N1 are in
  Kozai-Lidov resonance with their argument of pericenter oscillating
  around 90 deg, but Psamathe is not in this resonance.
  \label{fig:neso_group}}
\end{figure}

At Neptune, S/2021 N1 has a similar retrograde orbit as Psamathe and
Neso called the Neso group in Figure~\ref{fig:Allai2024JupSat}. We
find S/2021 N1 is in a Kozai-Lidov resonance, like Neso
(Figure~\ref{fig:kozai2}). The other member of the Neso Neptunian
satellite group, Psamathe, does not appear to be in the Kozai-Lidov
resonance (Figure~\ref{fig:neso_group}). The Neso dynamical group is a
strong dynamical grouping as both the semi-major axis and inclination
osculating orbital elements overlap for all three objects and they
have minimum dispersion velocities all below 100 m/s
(Tables~\ref{tab:neptune_osc} and~\ref{tab:dispvelocity}). Neso has
about $80 \%$ of the total volume of the group and the parent
satellite of the Neso group would have had a diameter of about 44 km
assuming an albedo of 0.1.

\subsection{Neptunian Sao Dynamical Group}
\begin{figure}
\plotone{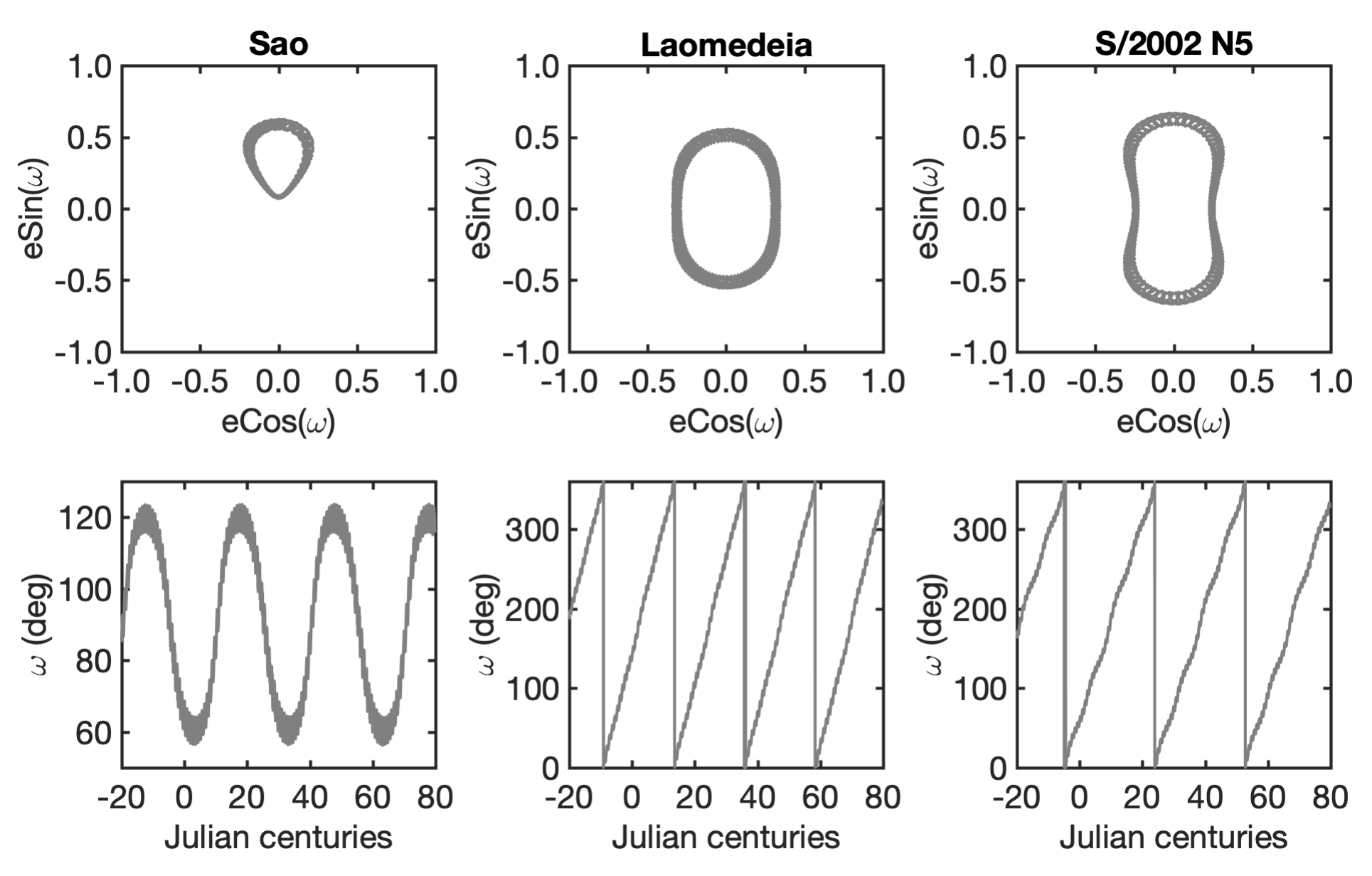}
\caption{Same as Figure~\ref{fig:kozai2} but for the Sao group of
  Neptune satellites. Sao is in the Kozai-Lidov resonance with its
  argument of pericenter oscillating around 90 deg, but other
  dynamical group members Laomedeia and S/2002 N5 are not in this
  resonance.
  \label{fig:sao_group}}
\end{figure}

S/2002 N5 has a similar prograde orbit to Sao and Laomedeia that is
labeled the Sao group in Figure~\ref{fig:Allai2024JupSat}.  The new
satellite S/2002 N5 currently has a semi-major axis and inclination
between the two brighter, and presumably larger satellites Sao and
Laomedeia. The Sao group's members all overlap in semi-major axis and
inclination osculating orbital elements (Table~\ref{tab:neptune_osc}).
They do have somewhat higher dispersion velocities than the other
dynamical group at Neptune (Table~\ref{tab:dispvelocity}), but most
are still below a velocity dispersion minimum of 100 m/s and small
enough that each object likely originated from the same parent body,
especially with the Kozai-Lidov resonance effecting some of the
satellites.  Sao is in the Kozai-Lidov resonance (Holman et al. 2004),
but is the only member of the Sao group to be in the Kozai-Lidov
resonance (Figure~\ref{fig:sao_group}).

The Sao group is unusual in that both the two largest members are of
similar brightness and likely size with newly discovered S/2002 N5
just a few tenths of a magnitude fainter and thus likely just a few km
smaller. The parent satellite of this group would have had an
effective diameter of about 38 km assuming an albedo of 0.1.

\section{Small Outer Satellite Size Distributions}

\begin{figure}
\plotone{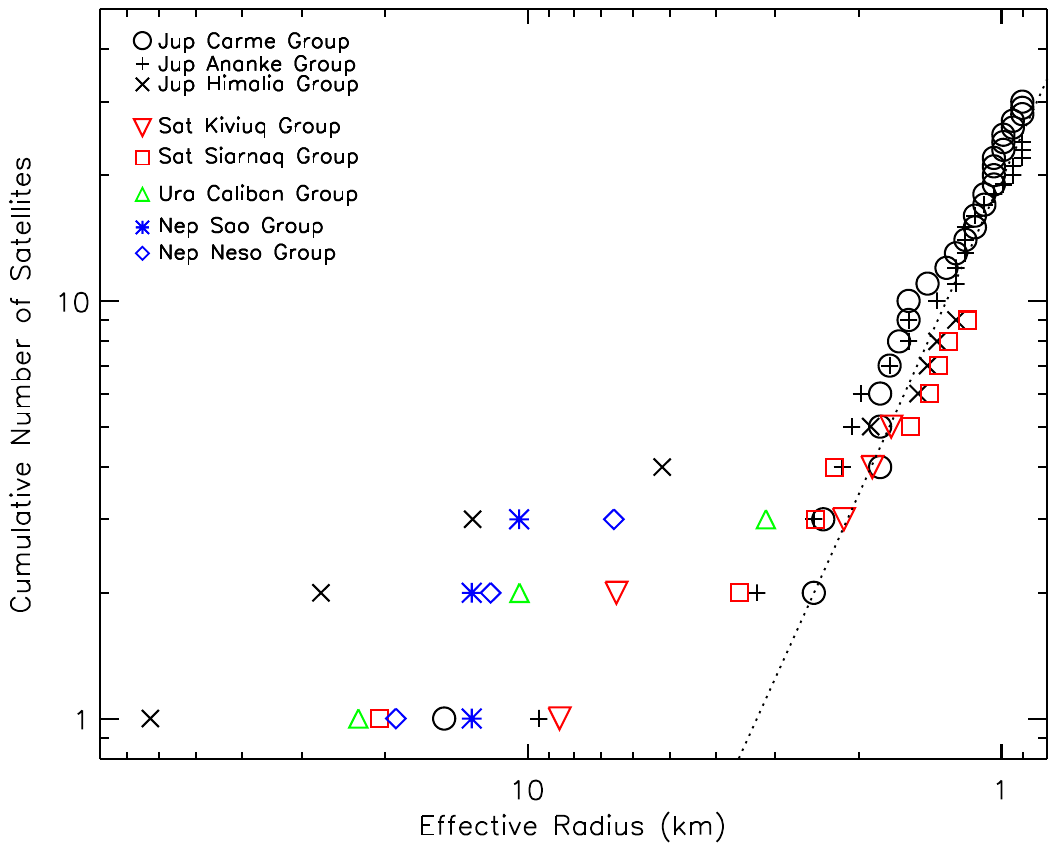}
\caption{The cumulative size distribution for the various
  well-determined outer irregular satellite dynamical families around
  the giant planets. Most dynamical satellite families have a few
  large to medium members and many more smaller members. A typical
  steep collisional size distribution is apparent for satellites
  smaller than about 5 km in size, where as larger satellites have a
  much shallower distribution. The dotted line shows the canonical
  Dohnanyi (1972) asteroid collisional power law of $q=-2.5$, which
  fits the smaller satellites of less than 5 km well. This indicates
  the smallest known satellites are consistent with a collisional origin.
  \label{fig:satsizeall}}
\end{figure}

In Figure~\ref{fig:satsizeall} we plot the cumulative size
distribution of the outer satellite dynamical families of the giant
planets that have tight orbital clustering and more than two known
members. Thus we do not use the Jupiter prograde Carpo group, which is
unusual in that it only has two small members and no large member. We
also do not use the Jupiter Pasiphae/Sinope group as it is not as
tightly confined in orbital space as the other Jupiter satellite
dynamical families, though it appears to have a similar size power law
as the other Jupiter families with one or two large members and many
more smaller members.  We further do not use the Saturn Gallic group
as it is not as tightly clustered as the Saturn Kiviuq and Siarnaq
groups. The Gallic group, if a true dynamical family, would be unusual
in that not only is it somewhat dispersed, but it has several large or
medium sized members and very few small members known (Sheppard et
al. 2023). It is unclear if there are Saturn retrograde dynamical
families or not, so we do not use any of these Saturn retrograde
satellites in our dynamical family size power law analysis.

It is seen in Figure~\ref{fig:satsizeall} that there is generally a
shallow size power law for the largest few satellites of a dynamical
family and then a steep power law for the smaller satellites of a
family, meaning many more small members (Sheppard \& Jewitt 2003;
Sheppard et al. 2006, 2018, 2023; Bottke et al. 2010; Alexandersen et
al. 2012; Nesvorny 2018; Ashton et al. 2020, 2021). The cumulative
number power-law size index is typically represented by $q$, where
$N(>r) \propto r^{q}$. The standard Dohnanyi et al. (1972) collisional
cumulative power law size frequency distribution index has a steep
slope of $q\sim -2.5$, which we plot as a dotted line in
Figure~\ref{fig:satsizeall}. We note that this steep slope, consistent
with a collisional size distribution, starts for the smallest outer
satellites around 5 km in diameter. This is consistent with an
interpretation that most satellites smaller than 5 km are products of
collisional disruption.

There is a significant increase in the number of satellites for a
family starting around a diameter of 5 km. At Uranus and Neptune we
have not yet efficiently surveyed for such small outer satellites, as
mentioned above being only complete to about 8 and 14 km for outer
satellites around these planets, respectively. Thus if the dynamical
groups are from a once larger parent satellite, many more smaller and
likely fainter satellites are expected to exist for the Caliban, Sao and
Neso dynamical groups at Uranus and Neptune.

Besides the large satellites Triton and Nereid, which have much closer
orbits to Neptune and may have been inner satellites as described in
the introduction, the only other known normal outer irregular Neptune
satellite is Halimede. Interestingly, Halimede, to date a lone member,
is of a similar diameter as the parent satellites of the Neso and Sao
Neptune dynamical Groups, being about 42 km in size assuming an albedo
of 0.1. Thus Neptune might of only had three original distant outer
irregular satellites, all of similar size: the Neso parent body, the
Sao parent body and Halimede. One might expect there to be several
lone or parent outer irregular satellites of Neptune in the 20 km size
range, as smaller objects should have been more numerous and thus more
likely to be captured as satellites. The lack of original $\sim 20$ km
sized Neptune outer satellites might be a sign that asteroids formed
big and were captured as satellites before significant break-up
occurred to the asteroids (Morbidelli et al. 2009; Sheppard et
al. 2010; Shankman et al. 2013; Johansen \& Lambrechts 2017). If true,
satellites at Neptune smaller than 40 to 50 km would likely only be
found as fragments of larger parent objects, like appears to be the
case for S/2021 N1 and S/2002 N5.

\section{Conclusions and Summary}

Through an extra ultra-deep pencil-beam survey of the space near
Uranus and Neptune we have imaged over a magnitude fainter than
previous surveys and found three new outer satellites.  S/2023 U1 is a
new retrograde Uranian satellite of 26.6 mags in the r-band,
corresponding to about 7 km in diameter assuming a ten percent
albedo. S/2021 N1 is a new retrograde Neptunian satellite that is 26.9
mags, corresponding to 14 km in diameter while S/2002 N5 is a prograde
satellite and is 25.9 mags corresponding to 23 km in diameter. We
detected, on multiple nights, all known outer satellites of Uranus and
Neptune during these observations in 2021, 2022 and 2023. All new
astrometry has been reported to the Minor Planet Center as some of
these satellites have not been observed since 2004.  These
observations nearly complete the outer satellite populations of Uranus
and Neptune to about 26.5 and 27 mags, corresponding to diameters of
about 8 and 14 km assuming albedos of 0.1, respectively.

All the new satellite discoveries have osculating orbital elements
that overlap significantly with two larger known satellites, as well
as have low dispersion velocities with those same known
satellites. This gives three members for each of these identified
dynamical families for the first time. At Uranus there is the
retrograde Caliban group with Caliban, Stephano and S/2023 U1 where
Stephano and S/2023 U1 overlap almost completely in osculating orbital
phase space with a minimum dispersion velocity of only 21 m/s. The
likely parent satellite of the Caliban group was about 45 km in
diameter. This Uranian dynamical grouping shows outer irregular
satellites of Uranus were likely broken apart over time and it is
possible the dust produced from these events could be the source of
the red material seen on the leading hemispheres of some the inner
larger Uranian satellites.

At Neptune there is the prograde Sao dynamical group with Sao,
Laomedeia and S/2002 N5, where all three members overlap in osculating
orbital phase space and have minimum dispersion velocities of less
than 80 m/s with each other. The Sao group parent satellite may have
had a diameter of about 38 km.  The Neptune retrograde Neso group
includes Neso, Psamathe and S/2021 N1, with the parent satellite about
44 km in diameter. We find that newly discovered S/2021 N1 is in a
Kozai-Lidov resonance as its argument of pericenter librates around 90
degrees.

The satellite dynamical families of the giant planets significantly
increase in the number of members for diameters less than about 5 km
in diameter. Steep size distribution slopes consistent with
collisional breakup of once larger parent satellites is seen for
diameters less than 5 km. For both the Uranus and Neptune dynamical
groups, we expect many more smaller satellites exist, but current
surveys have not gone deep enough to efficiently discover satellites
around these planets as small or smaller than 5 km in diameter. If
most asteroids formed bigger than $\sim 40-50$ km in size through
pebble accretion, this might explain the lack of $\sim 20$ km sized
parent outer satellites at Neptune. Captured outer satellites at
Neptune smaller than $\sim 40-50$ km may only be found as fragments of
larger parent objects, like appears to be the case for S/2021 N1 and
S/2002 N5.

\section*{Acknowledgments}

This paper includes data gathered with the 6.5 meter Magellan
Telescopes located at Las Campanas Observatory, Chile. This research
is based in part on data collected at the Subaru Telescope, which is
operated by the National Astronomical Observatory of Japan. We are
honored and grateful for the opportunity of observing the Universe
from Maunakea, which has the cultural, historical, and natural
significance in Hawaii. Based on observations obtained at the
international Gemini Observatory, a program of NSF’s NOIRLab, which is
managed by the Association of Universities for Research in Astronomy
(AURA) under a cooperative agreement with the National Science
Foundation on behalf of the Gemini Observatory partnership: the
National Science Foundation (United States), National Research Council
(Canada), Agencia Nacional de Investigaci\'{o}n y Desarrollo (Chile),
Ministerio de Ciencia, Tecnolog\'{i}a e Innovaci\'{o}n (Argentina),
Minist\'{e}rio da Ci\^{e}ncia, Tecnologia, Inova\c{c}\~{o}es e
Comunica\c{c}\~{o}es (Brazil), and Korea Astronomy and Space Science
Institute (Republic of Korea). Gemini program ID GN-2023B-DD-102.
Part of the research described here was carried out at the Jet
Propulsion Laboratory, California Institute of Technology, under
contract with the National Aeronautics and Space Administration
(80NM0018D0004). Based on observations collected at the European
Southern Observatory under ESO programme 110.259D.001.

\clearpage

\newpage

\clearpage



\startlongtable
\begin{deluxetable}{lccccc}
\tablenum{1}
\tablewidth{7.0 in}
\tablecaption{Fields Imaged Near Uranus and Neptune\label{tab:fields2024}}
\tablecolumns{6}
\tablehead{
\colhead{UT Date} & \colhead{Tel} & \colhead{Center}   & \colhead{RA and Dec}  &  \colhead{$\theta$}  &   \colhead{Limit} \\ \colhead{yyyy/mm/dd} & \colhead{} &  \colhead{} & \colhead{(hrs and degs)}  &  \colhead{(``)}  &  \colhead{(m$_{r}$)}}
\startdata
2021/09/03 & Mag & NeptWest     &  23:30:43 -04:09:10 & $0.73$ &  $26.6$   \\
2021/09/03 & Mag & NeptEast     &  23:32:39 -04:37:10 & $0.67$ &  $26.7$   \\
2021/09/07 & Sub & NeptCenter   &  23:31:15 -04:19:43 & $0.83$ &  $27.2$   \\
2021/09/08 & Sub & NeptCenter   &  23:31:09 -04:20:24 & $0.76$ &  $26.9$   \\
2021/09/08 & Sub & UranCenter   &  02:48:01 +15:42:20 & $0.73$ &  $26.9$   \\
2021/10/06 & Mag & S2002N5      &  23:27:52 -04:40:14 & $0.45$ &  $26.4$   \\
2021/10/06 & Mag & S2021N1      &  23:30:23 -04:41:33 & $0.45$ &  $26.9$   \\
2021/10/29 & Mag & UranSE       &  02:42:34 +15:08:46 & $0.9$  &  $26.3$   \\
2021/10/30 & Mag & UranNE       &  02:42:32 +15:21:46 & $0.53$ &  $26.8$   \\
2021/10/30 & Mag & UranSW       &  02:40:41 +15:08:03 & $0.62$ &  $26.7$   \\
2021/12/02 & Mag & UranNW       &  02:35:32 +14:59:00 & $0.68$ &  $26.6$   \\
2021/12/06 & Mag & UranSE       &  02:36:52 +14:36:19 & $0.84$ &  $26.5$   \\
2021/12/07 & Mag & Uran2SW      &  02:34:44 +14:27:27 & $0.77$ &  $26.7$   \\
2021/12/07 & Mag & S2002N5      &  23:24:46 -04:57:46 & $0.79$ &  $26.0$   \\
2022/10/15 & Mag & S2002N5      &  23:37:09 -03:53:01 & $0.59$ &  $26.6$   \\
2022/10/16 & Mag & S2002N5      &  23:37:15 -03:52:15 & $0.46$ &  $26.3$   \\
2022/11/16 & VLT & S2021N1      &  23:36:25 -04:11:38 & $0.81$ &  $27.2$   \\
2023/11/03 & Gem & S2021N1      &  23:45:28 -03:23:02 & $0.70$ &  $27.3$   \\
2023/11/04 & Mag & 21N1,02N5    &  23:45:01 -03:17:36 & $0.58$ &  $26.9$   \\
2023/11/04 & Mag & UranCenter   &  03:15:07 +17:43:36 & $0.67$ &  $26.9$   \\
2023/12/06 & Mag & S2023U1      &  03:09:44 +17:25:55 & $0.75$ &  $26.9$   \\
2023/12/13 & Mag & S2023U1      &  03:08:57 +17:26:00 & $0.78$ &  $26.8$   \\
\enddata
\tablenotetext{}{The telescopes (Tel) and  wide-field instruments used were HyperSuprime-Cam on Subaru (Sub) and IMACS on Magellan (Mag). Recovery was done with IMACS as well as GMOS on Gemini North (Gem) and FORS2 on the VLT. Center is the planet observed: Neptune (Nept) or Uranus (Uran) along with the position of the detector relative to the planet or centered on a new moon observed. The Subaru observations always had the planet near the center of the dectector. $\theta$ is the average seeing for the set of images and Limit is the
limiting magnitude in the r-band where we would have found at least
75\% of the satellites.  Under the basic survey information for each night
are the fields observed in J2000 coordinates for Right Ascension
(RA hh:mm:ss) and Declination (Dec dd:mm:ss).}
\end{deluxetable}



\newpage

\startlongtable
\begin{deluxetable}{lcrrrrr} 
\tablenum{2} 
\tablecaption{New Absolute Astrometry And Residuals For Outer Satellites Of Neptune\label{tab:residsnep}} 
\tablewidth{0pt}
\tablehead{ \colhead{Object} & \colhead{Site} & \colhead{Time} & \colhead{$\alpha$} & \colhead{$\delta$} & \colhead{res. $\alpha cos(\delta)$} & \colhead{res. $\delta$} \\
 \colhead{} & \colhead{} & \colhead{UTC} & \colhead{hh mm ss} & \colhead{deg mm sec} & \colhead{arcsecond} & \colhead{arcsecond}
}
\startdata
Halimede   & 568  &  2021 09 07.45001 & 23 31 03.576 & -04 12 28.66  &    0.127    & -0.131  \\
Halimede   & 568  &  2021 09 07.49866 & 23 31 03.277 & -04 12 30.49  &    0.042    & -0.072  \\
Halimede   & 568  &  2021 09 08.45824 & 23 30 57.596 & -04 13 07.75  &   -0.057    & -0.010  \\
Halimede   & 568  &  2021 09 08.48150 & 23 30 57.456 & -04 13 08.50  &   -0.047    &  0.145  \\
Halimede   & 568  &  2021 09 08.50087 & 23 30 57.330 & -04 13 09.28  &   -0.182    &  0.118  \\
Psamathe   & 568  &  2021 09 07.45442 & 23 33 18.749 & -04 47 16.73  &    0.127    &  0.295 \\  
Psamathe   & 568  &  2021 09 07.52961 & 23 33 18.281 & -04 47 20.08  &    0.047    & -0.053 \\
Psamathe   & 568  &  2021 09 08.45824 & 23 33 12.679 & -04 47 57.15  &    0.132    & 0.001 \\
Sao        & 568  &  2021 09 07.45885 & 23 32 07.372 & -04 26 26.41  &   0.219   & -0.011  \\  
Sao        & 568  &  2021 09 07.52961 & 23 32 06.925 & -04 26 29.29  &  -0.041   & -0.236  \\ 
Sao        & 568  &  2021 09 08.45824 & 23 32 01.388 & -04 27 03.84  &  -0.106   &  0.112  \\ 
Sao        & 568  &  2021 09 08.48150 & 23 32 01.259 & -04 27 04.88  &   0.085   & -0.054  \\ 
Sao        & 568  &  2021 09 08.50087 & 23 32 01.136 & -04 27 05.44  &   0.009   &  0.114  \\ 
Laomedeia  & 568  &  2021 09 07.45442 & 23 31 32.135 & -04 12 23.72  &  -0.024     &  -0.306  \\
Laomedeia  & 568  &  2021 09 07.49866 & 23 31 31.861 & -04 12 25.30  &   0.061     &  -0.081  \\
Laomedeia  & 568  &  2021 09 07.52961 & 23 31 31.667 & -04 12 26.48  &   0.081     &   0.001  \\
Laomedeia  & 568  &  2021 09 08.48150 & 23 31 25.759 & -04 13 05.36  &   0.062     &   0.013  \\
Laomedeia  & 568  &  2021 09 08.50087 & 23 31 25.640 & -04 13 06.31  &   0.118     &  -0.146  \\
Neso       & 568  &  2021 09 07.45001 & 23 33 11.178 & -04 52 56.16  &  -0.091     &   0.181  \\    
Neso       & 568  &  2021 09 07.46327 & 23 33 11.091 & -04 52 56.82  &  -0.165     &   0.055 \\  
Neso       & 568  &  2021 09 08.46987 & 23 33 04.979 & -04 53 37.52  &  -0.190     &   0.023 \\
Neso       & 568  &  2021 09 08.45824 & 23 33 05.047 & -04 53 36.95  &  -0.128     &  -0.078 \\ 
S/2002 N5 &304 & 2021 09 03.16979  &  23 31 19.070  &  -04 17 09.73  &  0.404   &   0.557 \\
S/2002 N5 &304 & 2021 09 03.20208  &  23 31 18.880  &  -04 17 11.12  &  0.388   &   0.338 \\
S/2002 N5 &304 & 2021 09 03.24528  &  23 31 18.610  &  -04 17 12.87  &  0.132   &   0.153 \\
S/2002 N5 &568 & 2021 09 07.45885  &  23 30 54.198  &  -04 19 47.18  &  0.084   &  -0.245 \\
S/2002 N5 &568 & 2021 09 07.52961  &  23 30 53.766  &  -04 19 49.69  & -0.081   &  -0.163 \\
S/2002 N5 &568 & 2021 09 08.45824  &  23 30 48.348  &  -04 20 23.69  & -0.084   &  -0.078 \\
S/2002 N5 &568 & 2021 09 08.46597  &  23 30 48.301  &  -04 20 23.76  & -0.097   &   0.135 \\
S/2002 N5 &568 & 2021 09 08.46987  &  23 30 48.265  &  -04 20 24.03  & -0.288   &   0.009 \\
S/2002 N5 &568 & 2021 09 08.48924  &  23 30 48.161  &  -04 20 24.88  & -0.114   &  -0.131 \\
S/2002 N5 &568 & 2021 09 08.49311  &  23 30 48.143  &  -04 20 25.12  & -0.038   &  -0.229 \\
S/2002 N5 &568 & 2021 09 08.50087  &  23 30 48.107  &  -04 20 25.19  &  0.116   &  -0.014 \\
S/2002 N5 &304 & 2021 10 06.05016  &  23 28 09.955  &  -04 36 45.38  &  0.064   &  -0.084 \\
S/2002 N5 &304 & 2021 10 06.05974  &  23 28 09.899  &  -04 36 45.73  &  0.006   &  -0.120 \\ 
S/2002 N5 &304 & 2021 12 07.05600  &  23 25 26.695  &  -04 52 02.60  & -0.341   &  -0.344 \\
S/2002 N5 &304 & 2022 10 15.06092  &  23 37 02.582  &  -03 47 17.28  &  0.011   &  -0.102 \\
S/2002 N5 &304 & 2022 10 15.13403  &  23 37 02.195  &  -03 47 19.68  & -0.090   &  -0.050 \\
S/2002 N5 &304 & 2022 10 16.06909  &  23 36 57.474  &  -03 47 50.71  & -0.040   &   0.055 \\
S/2002 N5 &304 & 2022 10 16.08325  &  23 36 57.398  &  -03 47 51.34  & -0.088   &  -0.106 \\
S/2002 N5 &304 & 2022 10 16.09739  &  23 36 57.341  &  -03 47 51.55  &  0.147   &   0.151 \\
S/2002 N5 &304 & 2023 11 04.04097  &  23 44 26.210  &  -03 11 35.83  & -0.306   &   0.153 \\
S/2002 N5 &304 & 2023 11 04.12662  &  23 44 25.922  &  -03 11 38.18  &  0.402   &  -0.132 \\
S/2021 N1 &568 & 2021 09 07.45001  &  23 33 34.149  &  -04 19 20.31  &   0.117  &   -0.170 \\
S/2021 N1 &568 & 2021 09 07.48980  &  23 33 33.890  &  -04 19 21.93  &  -0.134  &   -0.175 \\
S/2021 N1 &568 & 2021 09 07.52961  &  23 33 33.668  &  -04 19 23.51  &   0.167  &   -0.140 \\
S/2021 N1 &568 & 2021 09 08.45824  &  23 33 28.114  &  -04 20 01.13  &   0.020  &   -0.009 \\
S/2021 N1 &568 & 2021 09 08.50087  &  23 33 27.866  &  -04 20 02.84  &   0.204  &    0.014 \\
S/2021 N1 &304 & 2021 10 06.09365  &  23 30 44.708  &  -04 38 09.62  &  -0.020  &    0.129 \\
S/2021 N1 &304 & 2021 10 06.10324  &  23 30 44.654  &  -04 38 09.79  &  -0.017  &    0.307 \\
S/2021 N1 &304 & 2021 10 06.13223  &  23 30 44.428  &  -04 38 11.08  &  -0.947  &    0.068 \\
S/2021 N1 &309 & 2022 11 16.02415  &  23 36 28.290  &  -04 11 34.49  &  -0.047  &   -0.040 \\
S/2021 N1 &309 & 2022 11 16.12292  &  23 36 28.057  &  -04 11 35.74  &   0.019  &    0.027 \\
S/2021 N1 &568 & 2023 11 03.20694  &  23 45 28.539  &  -03 23 03.60  &  -0.075  &    0.102 \\
S/2021 N1 &568 & 2023 11 03.27194  &  23 45 28.281  &  -03 23 05.05  &  -0.028  &    0.216 \\
S/2021 N1 &304 & 2023 11 04.04097  &  23 45 25.308  &  -03 23 23.46  &   0.039  &   -0.167 \\
S/2021 N1 &304 & 2023 11 04.12639  &  23 45 24.975  &  -03 23 25.46  &   0.091  &   -0.162 \\
\enddata
\tablecomments{Residuals for absolute astrometry in Right Ascension ($\alpha$) and Declination ($\delta$). Results use the Neptune satellites nep104 orbit solutions from the JPL {\it Horizons} On-Line Solar System Data Service (Acton 1996) and from NASAs Navigation and Ancillary Information Facility (Giorgini et al. 1996) (\href{https://ssd.jpl.nasa.gov/sats}{ssd.jpl.nasa.gov/sats}). Site codes are 568 for Mauna Kea in Hawaii, 304 for Magellan at Las Campanas in Chile and 309 for the Very Large Telescope in Chile.}
\end{deluxetable}

\newpage

\startlongtable
\begin{deluxetable}{lcrrrrr} 
\tablenum{3} 
\tablecaption{New Absolute Astrometry and Residuals For The Outer Satellites Of Uranus \label{tab:residsura}} 
\tablewidth{0pt}
\tablehead{ \colhead{Object} & \colhead{Site} & \colhead{Time} & \colhead{$\alpha$} & \colhead{$\delta$} & \colhead{res. $\alpha cos(\delta)$} & \colhead{res. $\delta$} \\
 \colhead{} & \colhead{} & \colhead{UTC} & \colhead{hh mm ss} & \colhead{deg mm sec} & \colhead{arcsecond} & \colhead{arcsecond}
}
\startdata
Caliban  &   568 &    2021 09 08.50737& 02 47 37.867&+15 37 34.32      &   -0.048   & -0.111 \\
Caliban  &   568 &    2021 09 08.52656& 02 47 37.794&+15 37 33.88      &   -0.058   & -0.194 \\
Caliban  &   568 &    2021 09 08.54960& 02 47 37.705&+15 37 33.37      &   -0.086   & -0.272 \\
Caliban  &   304 &    2021 10 30.21467& 02 41 24.923&+15 08 11.58      &   -0.114   &  0.031 \\
Caliban  &   304 &    2023 11 04.14634& 03 15 36.150&+17 47 08.14      &   -0.107   & -0.180 \\
Caliban  &   304 &    2023 11 04.18390& 03 15 35.783&+17 47 06.93      &    0.073   & -0.123 \\
Caliban  &   304 &    2023 11 04.23291& 03 15 35.280&+17 47 05.28      &   -0.045   & -0.108 \\
Sycorax  &   568 &    2021 09 08.50737& 02 47 27.293&+15 36 26.14      &   0.009    &   0.047 \\
Sycorax  &   568 &    2021 09 08.52656& 02 47 27.214&+15 36 25.70      &  -0.033    &  -0.084 \\
Sycorax  &   568 &    2021 09 08.54960& 02 47 27.121&+15 36 25.40      &  -0.053    &  -0.009 \\
Sycorax  &   304 &    2021 10 30.17326& 02 41 03.834&+15 08 44.64      &  -0.061    &   0.004 \\
Sycorax  &   304 &    2021 10 30.21467& 02 41 03.426&+15 08 42.96      &  -0.068    &   0.055 \\
Sycorax  &   304 &    2021 10 30.25226& 02 41 03.057&+15 08 41.33      &  -0.055    &   0.003 \\
Prospero &   304 &    2021 10 30.23342& 02 40 02.339&+15 16 54.72     &    0.279   & -0.237 \\ 
Prospero &   304 &    2021 10 30.24475& 02 40 02.214&+15 16 54.20     &    0.072   & -0.266 \\
Prospero &   304 &    2021 10 30.25605& 02 40 02.105&+15 16 53.79     &    0.100   & -0.185 \\
Prospero &   304 &    2023 11 04.14634& 03 15 46.483&+17 38 52.11     &   -0.112   &  0.190 \\
Prospero &   304 &    2023 11 04.18013&03 15 46.143&+17 38 50.79     &   -0.017   &  0.245 \\
Prospero &   304 &    2023 12 06.11627& 03 10 26.561&+17 17 13.55     &   -0.063   &  0.075 \\
Prospero &   304 &    2023 12 06.18583& 03 10 25.915&+17 17 11.01     &    0.016   &  0.140 \\
Setebos  &   568 &    2021 09 08.51120& 02 48 27.552&+15 53 29.32      &   -0.150   &   0.112 \\
Setebos  &   568 &    2021 09 08.52656& 02 48 27.502&+15 53 29.05      &   -0.052   &   0.059 \\
Setebos  &   568 &    2021 09 08.54576& 02 48 27.424&+15 53 28.89      &   -0.160   &   0.172 \\
Setebos  &   304 &    2021 10 30.26006& 02 42 13.952&+15 26 28.06      &   -0.049   &   0.191 \\
Setebos  &   304 &    2021 10 30.33414& 02 42 13.233&+15 26 24.89      &   -0.058   &   0.185 \\
Stephano &   568 &    2021 09 08.50737& 02 47 36.365&+15 42 04.72     &   0.009     &   -0.050 \\
Stephano &   568 &    2021 09 08.52656& 02 47 36.287&+15 42 04.26     &   0.016     &   -0.108 \\
Stephano &   568 &    2021 09 08.54960& 02 47 36.195&+15 42 03.72     &   0.054     &   -0.162 \\
Stephano &   304 &    2021 10 30.17326& 02 41 06.654&+15 09 47.98     &  -0.130     &    0.251 \\
Stephano &   304 &    2021 10 30.21467& 02 41 06.238&+15 09 45.94     &  -0.179     &    0.181 \\
Stephano &   304 &    2021 10 30.25226& 02 41 05.866&+15 09 44.10     &  -0.141     &    0.136 \\
Trinculo &   568 &    2021 09 08.50737& 02 47 37.865&+15 41 40.78     &  -0.071   &  0.005     \\
Trinculo &   568 &    2021 09 08.52656& 02 47 37.785&+15 41 40.56     &  -0.060   &  0.168     \\
Trinculo &   568 &    2021 09 08.54960& 02 47 37.694&+15 41 39.93     &   0.032   &  0.001 \\
Trinculo &   304 &    2021 10 30.17326& 02 41 03.668&+15 10 30.31     &  -0.068   &  0.226     \\
Trinculo &   304 &    2021 10 30.21467& 02 41 03.232&+15 10 28.30     &  -0.383   &  0.123     \\
Trinculo &   304 &    2021 10 30.25226& 02 41 02.882&+15 10 26.48     &   0.016   &  0.040     \\
Francisco &  568  &   2021 09 08.50737& 02 48 17.928&+15 45 49.77   &    0.179   &  -0.356    \\
Francisco &  568  &   2021 09 08.52656& 02 48 17.861&+15 45 49.70   &    0.406   &  -0.140    \\
Francisco &  568  &   2021 09 08.54960& 02 48 17.757&+15 45 49.18   &    0.329   &  -0.313    \\
Francisco &  304  &   2023 11 04.14258& 03 15 28.107&+17 44 27.39   &   -0.005   &  -0.092    \\
Francisco &  304  &   2023 11 04.14634& 03 15 28.068&+17 44 27.26   &   -0.006   &  -0.090    \\
Francisco &  304  &   2023 12 06.06707& 03 10 01.555&+17 25 24.57   &   -0.357   &   0.257    \\
Francisco &  304  &   2023 12 06.07125& 03 10 01.518&+17 25 24.44   &   -0.306   &   0.267    \\
Margaret  &  568  &   2021 09 08.50737 &02 47 28.368&+15 19 24.09    &  -0.322    &  -0.097 \\
Margaret  &  568  &   2021 09 08.52656 &02 47 28.303&+15 19 23.76    &  -0.155    &  -0.096 \\
Margaret  &  568  &   2021 09 08.54960 &02 47 28.213&+15 19 23.47    &  -0.130    &   0.015 \\
Margaret  &  304  &   2021 12 07.10835 &02 35 15.415&+14 25 09.89    &   0.126    &   0.118 \\
Margaret  &  304  &   2021 12 07.12760 &02 35 15.266&+14 25 09.30    &   0.139    &   0.155 \\
Ferdinand &   568 &   2021 09 08.50737& 02 49 12.416&+15 53 39.87   &  -0.044   &  0.037  \\
Ferdinand &   568 &   2021 09 08.52656& 02 49 12.341&+15 53 39.61   &  -0.030   &  0.109  \\
Ferdinand &   568 &   2021 09 08.54960& 02 49 12.245&+15 53 39.05   &  -0.097   & -0.049  \\
Ferdinand &   304 &   2021 10 30.26006& 02 42 43.789&+15 24 03.30   &  -0.141   &  0.010  \\
Ferdinand &   304 &   2021 10 30.29707& 02 42 43.438&+15 24 01.76   &   0.149   &  0.152  \\
Ferdinand &   304 &   2021 10 30.33414& 02 42 43.059&+15 23 59.86   &   0.015   & -0.059  \\
S/2023 U1 & 568 & 2021 09 08.50737 & 02 47 38.560 & +15 44 01.37 &    0.013 &    0.196 \\
S/2023 U1 & 568 & 2021 09 08.53040 & 02 47 38.453 & +15 44 00.74 &   -0.292 &   -0.123 \\
S/2023 U1 & 304 & 2021 12 02.12835 & 02 36 24.703 & +14 56 07.10 &    0.242 &   -0.124 \\
S/2023 U1 & 304 & 2021 12 02.20226 & 02 36 24.096 & +14 56 04.33 &    0.159 &   -0.296 \\
S/2023 U1 & 304 & 2023 11 04.14258 & 03 15 15.924 & +17 49 32.61 &   -0.203 &    0.257 \\
S/2023 U1 & 304 & 2023 11 04.19892 & 03 15 15.383 & +17 49 30.43 &    0.014 &    0.111 \\
S/2023 U1 & 304 & 2023 11 04.24063 & 03 15 14.992 & +17 49 28.77 &    0.322 &   -0.033 \\
S/2023 U1 & 304 & 2023 12 06.06707 & 03 10 09.059 & +17 29 53.08 &   -0.311 &    0.050 \\
S/2023 U1 & 304 & 2023 12 06.18594 & 03 10 08.016 & +17 29 48.92 &    0.047 &    0.015 \\
S/2023 U1 & 304 & 2023 12 13.09425 & 03 09 09.764 & +17 25 58.03 &   -0.174 &   -0.054 \\
S/2023 U1 & 304 & 2023 12 13.20694 & 03 09 08.857 & +17 25 54.51 &    0.085 &    0.060 \\
\enddata
\tablecomments{Residuals for absolute astrometry in Right Ascension ($\alpha$) and Declination ($\delta$). Results use the Uranus satellites ura117 orbit solutions from the JPL {\it Horizons} On-Line Solar System Data Service (Acton 1996) and from NASAs Navigation and Ancillary Information Facility (Giorgini et al. 1996) (\href{https://ssd.jpl.nasa.gov/sats}{ssd.jpl.nasa.gov/sats}). Site codes are 568 for Subaru on Mauna Kea in Hawaii and 304 for Magellan at Las Campanas in Chile.}
\end{deluxetable}

\newpage



\begin{center}
\begin{deluxetable}{lcccccccccc}
\tablenum{4}
\tablewidth{6.5 in}
\tablecaption{New Satellites of Uranus and Neptune\label{tab:newmoons}}
\tablecolumns{11}
\tablehead{
\colhead{Name} & \colhead{$a$} & \colhead{$e$}  & \colhead{$i$} & \colhead{$\Omega$} & \colhead{$\omega$} & \colhead{$M$}  & \colhead{P} & \colhead{Dia}  & \colhead{$m_{r}$} & \colhead{$H$}  \\ \colhead{} & \colhead{(AU)}  & \colhead{} &\colhead{(deg)} &\colhead{(deg)} &\colhead{(deg)} & \colhead{(deg)}  & \colhead{(yrs)}  & \colhead{(km)}  & \colhead{(mag)} & \colhead{(mag)} }  
\startdata
S/2023 U1  &  0.0533  &  0.187  &  141.89  &  265.84  &  142.16 & 24.08  & 1.86  & 7   & 26.6 & 13.7  \\
S/2002 N5  &  0.1562  &  0.548  &  42.133  &  274.15  &  62.73  & 105.52 & 8.60  & 23  & 25.9 & 11.2  \\
S/2021 N1  &  0.338   &  0.441  &  134.5   &  264.1   &  100.8  & 86.4   & 27.4  & 14  & 26.9 & 12.1 \\
\enddata
\tablenotetext{}{Quantities are the current new satellite orbits from the Minor Planet Electronic Circulars (MPECs) on the discovery of the new satellites published by the  Minor Planet Center (Sheppard et al. 2024a,2024b,2024c). Columns are the semi-major axis ($a$), eccentricity ($e$), inclination ($i$), longitude of the ascending node ($\Omega$), argument of perihelion ($\omega$), and Mean Anomaly ($M$) for Epoch 2024 Mar 31.0 with significant digits showing the size of uncertainty. Diameter (Dia) estimates assume a moderate albedo of 0.10. The r-band magnitude ($m_{r}$) is the opposition magnitude.}
\end{deluxetable}
\end{center}




\newpage

\begin{deluxetable}{clccc} 
\tablenum{5} 
\tablecaption{Mean Osculating Orbital Elements For Outer Satellites Of Uranus\label{tab:uranus_osc}} 
\tablewidth{0pt}
\tablehead{ \colhead{Satellite} & \colhead{Osc. a (km)} & \colhead{Osc. e} & \colhead{Osc. i (deg)} &\colhead{Osc. P (d)}}
\startdata
Francisco& $4275700_{-700}^{+700}$           &	  $0.14_{-0.05}^{+0.05}$  &  $147_{-1}^{+2}$ & $267_{-1}^{+1}$    \\
Caliban  & $7167000_{-5400}^{+5400}$         &    $0.20_{-0.13}^{+0.12}$      &  $141_{-2}^{+3}$  & $580_{-1}^{+1}$	 \\
Stephano & $7951400_{-8000}^{+8600}$         &	  $0.23_{-0.11}^{+0.11}$   &  $144_{-3}^{+3}$  & $677_{-1}^{+1}$   \\
S/2023 U1& $7976600_{-8600}^{+8600}$         &	  $0.25_{-0.11}^{+0.11}$   &  $144_{-3}^{+3}$  & $681_{-1}^{+1}$   \\
Trinculo & $8502600_{-10600}^{+12200}$       &	  $0.22_{-0.02}^{+0.02}$   &  $167_{-1}^{+1}$  & $749_{-1}^{+2}$    \\
Sycorax  & $12193200_{-51900}^{+65200}$      &	  $0.52_{-0.08}^{+0.08}$   &  $157_{-6}^{+6}$  & $1286_{-8}^{+10}$  \\
Margaret & $14425000_{-106500}^{+113500}$    &	  $0.64_{-0.23}^{+0.25}$   &  $61_{-15}^{+8}$  & $1655_{-18}^{+20}$  \\
Prospero & $16221000_{-146200}^{+183600}$    &	  $0.44_{-0.13}^{+0.14}$   &  $149_{-6}^{+6}$  & $1974_{-27}^{+34}$ \\
Setebos  & $17519800_{-221300}^{+278300}$    &	  $0.58_{-0.12}^{+0.13}$   &  $154_{-8}^{+8}$  & $2215_{-42}^{+53}$  \\
Ferdinand& $20421400_{-338500}^{+455900}$    &	  $0.40_{-0.09}^{+0.09}$   &  $169_{-2}^{+3}$  & $2788_{-69}^{+94}$   \\
\enddata
\tablecomments{Osculating orbital elements, {\it a}, {\it e}, and {\it i}, represent the mean and the extreme osculating values obtained from 10,000 years of integrated orbits. The elements were generated with respect to the ecliptic pole RA=270.00 and Dec=66.56 degrees.}
\end{deluxetable}


\newpage

\begin{deluxetable}{clccc} 
\tablenum{6} 
\tablecaption{Mean Osculating Orbital Elements For Outer satellites Of Neptune\label{tab:neptune_osc}} 
\tablewidth{0pt}
\tablehead{ \colhead{Satellite} & \colhead{Osc. a (km)} & \colhead{Osc. e} & \colhead{Osc. i (deg)} & \colhead{Osc. P (d)} \\
}
\startdata
Halimede   & $16590500_{-36700}^{+42000}$    & $0.52_{-0.33}^{+0.38}$   & $120_{-9}^{+25}$    & $1879_{-6}^{+7}$  \\
Sao        & $22239900_{-116900}^{+128200}$   & $0.30_{-0.23}^{+0.33}$   & $50_{-11}^{+5}$    & $2917_{-23}^{+25}$    \\
S/2002 N5  & $23414700_{-151700}^{+172800}$   & $0.43_{-0.21}^{+0.24}$   & $46_{-10}^{+8}$    & $3151_{-31}^{+35}$ \\
Laomedeia  & $23499900_{-152000}^{+186000}$   & $0.42_{-0.13}^{+0.14}$   & $37_{-7}^{+7}$     & $3168_{-31}^{+38}$     \\
Psamathe   & $47646600_{-1983400}^{+2522400}$ & $0.41_{-0.34}^{+0.47}$   & $128_{-11}^{+20}$   & $9149_{-566}^{+735}$ \\
Neso       & $49897800_{-2242500}^{+2912900}$ & $0.46_{-0.32}^{+0.42}$   & $128_{-11}^{+17}$   & $9805_{-655}^{+870}$   \\
S/2021 N1  & $50700200_{-2253100}^{+2977200}$ & $0.50_{-0.18}^{+0.21}$   & $135_{-7}^{+7}$    & $10043_{-664}^{+713}$ \\
\enddata
\tablecomments{Osculating orbital elements, {\it a}, {\it e}, and {\it i}, represent the mean and the extreme osculating values obtained from 10,000 years of integrated orbits. The elements were generated with respect to the ecliptic pole, RA=270.00 and Dec=66.56 degs.}
\end{deluxetable}


\newpage

\begin{deluxetable}{lcc} 
\tablenum{7} 
\tablecaption{Dispersion Velocity Between Satellite Group Members \label{tab:dispvelocity}} 
\tablewidth{0pt}
\tablehead{ \colhead{Objects} & \colhead{$\delta V$} & \colhead{$\delta V_{min}$} \\ \colhead{} & \colhead{($m/s$)} & \colhead{$(m/s)$}
}
\startdata
\bf{Uranian Caliban Group} &   & \\
Caliban-Stephano    &  $257\pm47$   & 210 \\
Caliban-S/2023U1    &  $267\pm51$   & 216 \\
Stephano-S/2023U1   &  $66\pm45$    & 21 \\
\bf{Neptunian Sao Group} &   & \\
Sao-Laomedeia       & $200\pm62$    & 138 \\
Sao-S/2002N5        & $143\pm76$    & 67 \\
Laomedeia-S/2002N5  & $145\pm47$    & 98 \\
\bf{Neptunian Neso Group} &   & \\
Psamathe-Neso       & $120\pm79$    & 41 \\
Psamathe-S/2021N1   & $123\pm47$    & 76 \\
Neso-S/2021N1       & $107\pm46$    & 61 \\
\enddata
\tablecomments{$\delta V$ is the dispersion velocity between members in m/s while $\delta V_{min}$ is the minimal dispersion velocity. The dispersion velocity and its range were calculated over 30,000 years of orbital simulation with a data step of 100 days.}
\end{deluxetable}


\newpage

\newpage

\begin{references}

\reference{Act96} Acton, C. 1996, P\&SS, 44, 65.
  
\reference{Agn06} Agnor, C. \& Hamilton, D. 2006, Nature, 441, 192.

\reference{Ale12} Alexandersen, M., Gladman, B., Veillet, C., Jacobson, R., Brozovic, M. \& Rousselot, P. 2012, AJ, 144, 21.

\reference{Ash20} Ashton, E., Beaudoin, M. \& Gladman, B. 2020, PSJ, 1, 52.

\reference{Ash21} Ashton, E., Gladman, B. \& Beaudoin, M. 2021, PSJ, 2, 158.

\reference{Ash22} Ashton, E., Gladman, B., Beaudoin, M., Alexandersen, M. \& Petit, J. 2022, PSJ, 3, 107.

\reference{Bat20} Batygin, K. \& Morbidelli, A. 2020, ApJ, 894, 143.

\reference{Bea07} Beauge, C. \& Nesvorny, D. 2007, AJ, 133, 2537.

\reference{Bot10} Bottke, W., Nesvorny, D., Vokrouhlicky, D. \& Morbidelli, A. 2010, AJ, 139, 994.

\reference{Bot23a} Bottke, W., Marschall, R., Nesvorny, D. \& Vokrouhlicky, D. 2023, SSR, 219, 83.

\reference{Bot23b} Bottke, W., Vokrouhlicky, D., Marschall, R., Nesvorny, D., Morbidelli, A., Deienno, R., Marchi, S., Dones, L. \& Levison, H. 2023, PSJ, 4, 168.

\reference{Bro11} Brozovic, M., Jacobson, R. \& Sheppard, S. 2011, AJ, 141, 135.

\reference{Bro17} Brozovic, M. \& Jacobson, R. 2017, AJ, 153, 147.

\reference{Bro22} Brozovic, M. \& Jacobson, R. 2022, AJ, 163, 241.

\reference{Can02} Canup, R. \& Ward, W. 2002, AJ, 124, 3404.

\reference{Car02} Carruba, V., Burns, J., Nicholson, P. \& Gladman, B. 2002, Icarus, 158, 434.

\reference{Car04} Carruba, V., Nesvorny, D., Burns, J., Cuk, M. \& Tsiganis, K. 2004, AJ, 128, 1899.

\reference{Car16} Carruba, V. \& Nesvorny, D. 2016, MNRAS, 457, 1332.

\reference{Cha18} Charnoz, S., Canup, R., Crida, A. \& Dones, L. 2018. The Origin of Planetary Ring Systems. Planetary Ring Systems. Properties, Structure, and Evolution, Matthew S. Tiscareno and Carl D. Murray (eds.) ISBN: 9781316286791, 2018., Cambridge University Press, p.517-538.

\reference{Car18} Cartwright, R., Emery, J., Pinilla-Alonso, N., Lucas, M., Rivkin, A. \& Trilling, D. 2018, Icarus, 314, 210.

\reference{Car21} Cartwright, R., Beddingfield, C., Nordheim, T. et al. 2021, PSJ, 2, 120.

\reference{Car23} Cartwright, R., DeColibus, R., Castillo-Rogez, J., Beddingfield, C., Grundy, W. \& Nordheim, T. 2023, PSJ, 4, 42.

\reference{Col71} Colombo, G. \& Franklin, F. 1971, Icarus, 15, 186.

\reference{Coh22} Cohen, I., Beddingfield, C., Chancia, R. et al. 2022, PSJ, 3, 58.

\reference{Cuk04} Cuk, M. \& Burns, J. 2004, Icarus, 167, 369.

\reference{Cuk06} Cuk, M. \& Gladman, B. 2006, Icarus, 183, 362.

\reference{Cuk20a} Cuk, M., El Moutamid, M. \& Tiscareno, M. 2020a, PSJ, 1, 22.

\reference{Cuk20b} Cuk, M., French, R., Showalter, M., Tiscareno, M. \& El Moutamid, M. 2022, AJ, 164, 38.

\reference{Den19} Denk, T. \& Mottola, S. 2019, Icarus, 322, 80.

\reference{Doh72} Dohnanyi, J. 1972, Icarus, 17, 1.

\reference{Don11} Donnison, J. 2011, MNRAS, 415, 470.

\reference{Dre11} Dressler, A., Bigelow, B., Hare, T., et al. 2011, PASP, 123, 288.

\reference{Dur07} Durda, D., Bottke, W., Nesvorny, D., Enke, B., Merline, W., Asphaug, E. \& Richardson, D. 2007, Icarus, 186, 498.

\reference{Far17} Farkas-Takacs, A., Kiss, C., Pal, A. et al. 2017, AJ, 154, 119.

\reference{Fre15} French, R., Dawson, R. \& Showalter, M. 2015, AJ, 149, 142.

\reference{Fro11} Frouard, J., Vienne, A. \& Fouchard, M. A\&A, 2011, 532, 44.

\reference{Fur18} Furusawa, H. et al. 2018, PASJ, 70, S3.

\reference{Gas13} Gaspar, H., Winter, O. \& Vieira Neto, E. 2013, MNRAS, 433, 36.

\reference{Gio96} Giorgini, J., Yeomans, D., Chamberlin, A. et al. 1996, AAS/DPS Meeting Abstracts, 28, 25.04.

\reference{Gla01} Gladman, B., Kavelaars, J., Holman, M., et al. 2001, Nature, 412, 163.

\reference{Gom24} Gomes, R. \& Morbidelli, A. 2024, Icarus, 420, 116142.

\reference{Gra18} Graykowski, A. \& Jewitt, D. 2018, AJ, 155, 184.

\reference{Ham97} Hamilton, D. \& Krivov, A. 1997, Icarus, 128, 241.

\reference{Hol04} Holman, M., Kavelaars, J., Grav, T. et al. 2004, Nature, 430, 865.

\reference{Hol18} Holt, T., Brown, A., Nesvorn{\'y}, D., Horner, J., \& Carter, B. 2018, ApJ, 859, 97.

\reference{Hoo04} Hook, I., Jorgensen, I., Allington-Smith, J., Davies, R., Metcalfe, N., Murowinski, R. \& Crampton, D. 2004, PASP, 116, 425.

\reference{Jac12} Jacobson, R., Brozovic, M., Gladman, B., Alexandersen, M., Nicholson, P. \& Veillet, C. 2012, AJ, 144, 132.

\reference{Jac22} Jacobson, R., Brozovic, M., Mastrodemos, N., Riedel, J. \& Sheppard, S. 2022, AJ, 164, 240

\reference{Jew05} Jewitt, D. \& Sheppard, S. 2005, SSR, 116, 441.

\reference{Jew07} Jewitt, D. \& Haghighipour, N. 2007, ARAA, 45, 261.

\reference{Joh17} Johansen, A. \& Lambrechts, M. 2017, AREPS, 45, 359.

\reference{Kan23} Kane, S. \& Li, Z. 2023, PSJ, 4, 216.

\reference{Kav04} Kavelaars, J., Holman, M., Grav, T., Millisavljevic, D., Fraser, W., Gladman, B., Petit, J., Rousselot, P., Mousis, O. \& Nicholson, P. 2004, Icarus, 169, 474.

\reference{Kaw18} Kawanomoto, S. et al. 2018, PASJ, 70, 66.

\reference{Koc11} Koch, E. \& Hansen, B. 2011, MNRAS, 416, 1274.

\reference{Kom18} Komiyama, Y. et al. 2018, PASJ, 70, S2.

\reference{Li17} Li, D. \& Christou, A. 2017, AJ, 154, 209.

\reference{Li18} Li, D. \& Christou, A. 2018, Icarus, 310, 77.

\reference{Lix20} Li, D. \& Christou, A. 2020, AJ, 159, 184.

\reference{Lix20b} Li, D., Johansen, A., Mustill, A., Davies, M. \& Christou, A. 2020, AA, 638, A139.

\reference{Mic15} Michel, P., Richardson, D., Durda, D., Jutzi,
M., Asphaug, E., 2015.  Collisional Formation and Modeling of Asteroid
Families. Asteroids IV, Patrick Michel, Francesca E. DeMeo, and
William F. Bottke (eds.), University of Arizona Press, Tucson, 895
pp. ISBN: 978- 0-816-53213-1, 2015., p.341-354.

\reference{Miy18} Miyazaki, S. et al. 2018, PASJ, 70, S1.

\reference{Mor09} Morbidelli, A., Bottke, W., Nesvorny, D. \& Levison, H. 2009, Icarus, 204, 558.

\reference{Nes03} Nesvorny, D., Alvarellos, J., Dones, L., \& Levison, H. 2003, AJ, 126, 398.

\reference{Nes04} Nesvorny, D., Beauge, C. \& Dones, L. 2004, AJ, 127, 1768.

\reference{Nes07} Nesvorny, D., Vokrouhlicky, D. \& Morbidelli, A. 2007, AJ, 133, 1962.

\reference{Nes14} Nesvorny, D., Vokrouhlicky, D. \& Deienno, R. 2014, ApJ, 784, 22.

\reference{Nes15} Nesvorný, D., Brož, M., Carruba, V.,
2015. Identification and Dynamical Properties of Asteroid
Families. Asteroids IV, Patrick Michel, Francesca E. DeMeo, and
William F. Bottke (eds.), University of Arizona Press, Tucson, 895
pp. ISBN: 978-0-816-53213-1, 2015., p.297-321.

\reference{Nes18} Nesvorny, D. 2018, ARAA, 56, 137.

\reference{Nev19} Neveu, M. \& Rhoden, A. 2019, NatAs, 3, 543.

\reference{Nic08} Nicholson, P., Cuk, M., Sheppard, S., Nesvorny, D. \& Johnson, T. 2008. Irregular satellites of the giant planets. in: The Solar System Beyond Neptune, eds. M. Barucci, H. Boehnhardt, D. Cruikshank and A. Morbidelli, (The University of Arizona Press; Tucson) pp. 411-424.

\reference{Nog11} Nogueira, E., Brasser, R., \& Gomes, R. 2011, Icarus, 214, 113.

\reference{Par08} Parisi, M., Carraro, G., Maris, M. \& Brunini, A. 2008, AA, 482, 657.

\reference{Par21} Park, R., Folkner, W., Williams, J. \& Boggs, D. 2021, AJ, 161, 105.

\reference{Phi10} Philpott, C., Hamilton, D. \& Agnor, C. 2010, Icarus, 208, 824.

\reference{Ruf17} Rufu, R. \& Canup, R. 2017, AJ, 154, 208.

\reference{Ruf22} Rufu, R. \& Canup, R. 2022, ApJ, 928, 123.

\reference{Sal17} Salmon, J. \& Canup, R. 2017, ApJ, 836, 109.

\reference{Sha13} Shankman, C., Gladman, B., Kaib, N., Kavelaars, J. \& Petit, J. 2013, ApJ, 764, L2.

\reference{Sha23} Sharkey, B., Reddy, V., Kuhn, O., Sanchez, J. \& Bottke, W. 2023, PSJ, 4, 223.

\reference{She08} Shen, Y. \& Tremaine, S. 2008, AJ, 136, 2453.

\reference{She03} Sheppard, S. \& Jewitt, D. 2003, Nature, 423, 261.

\reference{She05} Sheppard, S., Jewitt, D. \& Kleyna, J. 2005, AJ, 129, 518.

\reference{She06} Sheppard, S., Jewitt, D. \& Kleyna, J. 2006, AJ, 132, 171.
  

\reference{She10} Sheppard, S. \& Trujillo, C. 2010, ApJ, 723, L233.

\reference{She16} Sheppard, S. \& Trujillo, C. 2016, AJ, 152, 221.

\reference{She18} Sheppard, S., Tholen, D., \& Trujillo, C. 2018, RNAAS, 7, 100.

\reference{She23} Sheppard, S., Tholen, D., Alexandersen, M. \& Trujillo, C. 2023, RNAAS, 7, 100.

\reference{She24a} Sheppard, S., Tholen, D., Trujillo, C., Lykawka, P. S., Brozovic, M. \& Jacobson, R.  2024a, MPEC 2024-D112.

\reference{She24b} Sheppard, S. et al. 2024b, MPEC 2024-D114.

\reference{She24c} Sheppard, S., Tholen, D., Trujillo, C., Brozovic, M. \& Jacobson, R. 2024c, MPEC 2024-D113.

\reference{Sho19} Showalter, M., de Pater, I., Lissauer, J. \& French, R. 2019, Nature, 566, 350.

\reference{Tam13} Tamayo, D., Burns, J. \& Hamilton, D. 2013, Icarus, 226, 655.

\reference{Vok08} Vokrouhlicky, D., Nesvorny, D. \& Levison, H. 2008, AJ, 136, 1463.

\end{references}
\end{document}